\PassOptionsToPackage{xcdraw,table}{xcolor}
\documentclass[acmsmall]{acmart}

\AtBeginDocument{%
  \providecommand\BibTeX{{%
    \normalfont B\kern-0.5em{\scshape i\kern-0.25em b}\kern-0.8em\TeX}}}

\acmBooktitle{Woodstock '18: ACM Symposium on Neural Gaze Detection,
  June 03--05, 2018, Woodstock, NY}
\setcopyright{acmlicensed}
\acmJournal{PACMHCI}
\acmYear{2021} \acmVolume{5} \acmNumber{CSCW1} \acmArticle{188} \acmMonth{4} \acmPrice{15.00}\acmDOI{10.1145/3449287}

\usepackage{hhline}
\usepackage{subcaption}
\usepackage{multirow}
\usepackage{makecell}

\begin{document}

\received{October 2020}
\received[revised]{January 2021}
\received[accepted]{January 2021}


\title[To Trust or to Think]{To Trust or to Think: Cognitive Forcing Functions Can Reduce Overreliance on AI in AI-assisted Decision-making}

\author{Zana Bu\c cinca}
\email{zbucinca@seas.harvard.edu}
\affiliation{%
  \institution{Harvard University}
  \streetaddress{33 Oxford St.}
  \city{Cambridge}
  \state{MA}
  \postcode{02138}
  \country{USA}
}

\author{Maja Barbara Malaya}
\email{217723@edu.p.lodz.pl}
\affiliation{%
  \institution{Lodz University of Technology}
  \streetaddress{ul. Stefana Żeromskiego 116}
  \postcode{90-924}
  \city{Łódź}
  \country{Poland}
}

\author{Krzysztof Z. Gajos}
\email{kgajos@eecs.harvard.edu}
\affiliation{%
  \institution{Harvard University}
  \streetaddress{33 Oxford St.}
  \city{Cambridge}
  \state{MA}
  \postcode{02138}
  \country{USA}
}

\newcommand{\paricipantnumber}{199 }

\begin{abstract}

People supported by AI-powered decision support tools frequently overrely on the AI: they accept an AI's suggestion even when that suggestion is wrong. Adding explanations to the AI decisions does not appear to reduce the overreliance and some studies suggest that it might even increase it.
Informed by the dual-process theory of cognition, we posit that people rarely engage analytically with each individual AI recommendation and explanation, and instead develop general heuristics about whether and when to follow the AI suggestions. Building on prior research on medical decision-making, we designed three cognitive forcing interventions to compel people to engage more thoughtfully with the AI-generated explanations.
We conducted an experiment (N=199), in which we compared our three cognitive forcing designs to two simple explainable AI approaches and to a no-AI baseline. The results demonstrate that cognitive forcing significantly reduced overreliance compared to the simple explainable AI approaches. However, there was a trade-off: people assigned the least favorable subjective ratings to the designs that reduced the overreliance the most. To audit our work for intervention-generated inequalities, we investigated whether our interventions benefited equally people with different levels of Need for Cognition (i.e., motivation to engage in effortful mental activities). Our results show that, on average, cognitive forcing interventions benefited participants higher in Need for Cognition more.
Our research suggests that human cognitive motivation moderates the effectiveness of explainable AI solutions. 

\end{abstract}



\maketitle

\section{Introduction}

From loan approval to disease diagnosis, humans are increasingly being assisted by artificially intelligent (AI) systems in decision-making tasks. By combining two types of intelligence, these emerging sociotechnical systems (i.e., human+AI teams) were expected to perform better than either people or AIs alone~\cite{kamar2012combining, kamar2016directions}. Recent studies, however, show that although human+AI teams typically outperform people working alone, their performance is usually inferior to the AI's~\cite{bucinca20:proxy,jacobs2021how,lai2019human,bussone2015role, bansal2020does,green2019principles,vaccaro2019effects}. There is evidence that instead of combining their own insights with suggestions generated by the computational models, people frequently overrely on the AI, following its suggestions even when those suggestions are wrong and the person would have made a better choice on their own~\cite{jacobs2021how,lai2019human,bussone2015role}.

Explainable AI (XAI), an approach where AI's recommendations are accompanied by explanations or rationales, was intended to address the problem of overreliance: By giving people an insight into how the machine arrived at its recommendations, the explanations were supposed to help them identify the situations in which AI's reasoning was incorrect and the suggestion should be rejected. However, evidence suggests that explainable systems, also, have not had substantial success in reducing human overreliance on the AIs: when the AI suggests incorrect or suboptimal solutions, people still on average make poorer final decisions than they would have without AI's assistance~\cite{jacobs2021how,lai2020why,zhang2020effect, bansal2020does,bucinca20:proxy}.

We posit that the dual-process theory provides a useful lens through which to understand the reasons why the explanations do not eliminate human overreliance on AIs. According to the dual-process theory, humans mostly operate on System 1 thinking, which employs heuristics and shortcuts when making decisions~\cite{wason1974dual, kahneman2011thinking}. Because most of the daily decisions are successfully accomplished by these heuristics, analytical thinking (i.e., System 2) is triggered rarely, as it is slower and costlier in terms of effort. Yet, System 1 thinking leaves us vulnerable to cognitive biases that can result in incorrect or suboptimal decisions~\cite{kahneman2011thinking}.
We do not want to suggest, however, that System 1 thinking is always bad or inappropriate. Indeed, successful use of pattern matching and heuristics that arise from extensive experience are important components of expertise~\cite{moulton2007slowing}. But even experts can fall prey to cognitive biases~\cite{lambe2016dual} so a judicious combination of both processes is needed.

In the context of explainable AI, the implicit assumption behind the design of most systems is that people will engage analytically with each explanation and will use the content of these explanations to identify which of the AI's suggestions are plausible and which appear to be based on faulty reasoning. Because evaluating every explanation requires substantial cognitive effort, which humans are averse to~\cite{kool18:mental}, this assumption is likely incorrect. Instead, people appear to develop heuristics about the competence of the AI partner overall. Indeed, some studies demonstrate that explanations are interpreted as a general signal of competence---rather than being evaluated individually for their content---and just by their presence can increase the trust in and overreliance on the AI~\cite{bansal2020does}. 

We argue that to reduce human overreliance on the AI and improve performance, we need to not only develop effective explanation techniques, but also ways to increase people's cognitive motivation for engaging analytically with the explanations.

A number of approaches have been explored in other domains for engaging people in more analytical thinking to reduce the impact of cognitive biases on decision-making. One of the most promising appears to be the \emph{cognitive forcing functions}---interventions that are applied at the decision-making time to disrupt heuristic reasoning and thus cause the person to engage in analytical thinking~\cite{lambe2016dual}. Examples of cognitive forcing functions include checklists, diagnostic time-outs, or asking the person to explicitly rule out an alternative. Bringing these insights into AI-assisted decision-making, we hypothesized that adding cognitive forcing functions to existing explainable AI designs would help reduce human overreliance on the AI. An important aspect of the design of forcing functions in this context, however, is the usability and the acceptability of these interventions. We hypothesized that stricter interventions will push people to think harder, but will be found complex and less usable.


We conducted an experiment with \paricipantnumber  participants on Amazon Mechanical Turk, in which we compared three cognitive forcing designs to two simple explainable AI approaches, and to a no-AI baseline. Our results demonstrate that cognitive forcing functions significantly reduced overreliance on the AI compared to the simple explainable AI approaches. However, cognitive forcing functions did not completely eliminate overreliance on the AI: even in cognitive forcing function conditions, participants were prone to rely on incorrect AI predictions for instances where they would have made a better decision without AI assistance.

As hypothesized, we also observed a trade-off between the acceptability of the designs and their effectiveness at reducing the overreliance on the AI. Specifically,
people over-relied less on the AI when exposed to the conditions that they found more difficult, preferred less, and trusted less.

We also audited our work for intervention-generated inequalities~\cite{veinot2018good}. Because prior work suggested that people with high Need for Cognition (NFC, a stable personality trait that captures one's motivation to engage in effortful mental activities) tend to benefit more from complex user interface features~\cite{carenini01:analysis,gajos17:influence,ghai2020explainable}, we did so by disaggregating our results by NFC level. We found that cognitive forcing functions benefited the more advantaged group---people with high NFC---the most.

In summary, we made two contributions in this paper:

First, we introduced \emph{cognitive forcing functions} as interaction design interventions for human+AI decision-making and demonstrated their potential to reduce overreliance on the AI. Our study also shows that there exists a trade-off between the effectiveness and the acceptability of interventions that cause people to exert more cognitive effort. Our research demonstrates that the effectiveness of human+AI collaborations on decision-making tasks depends on human factors, such as cognitive motivation, that go beyond the choice of the right form and content of AI-generated explanations. Hence, in addition to developing explanation techniques and tuning explanation attributes such as soundness and completeness~\cite{kulesza2013too}, or faithfulness, sensitivity, and complexity~\cite{bhatt2020evaluating}, more thought should be put into the design of the interaction to ensure that people will make effective use of the AI-generated recommendations and explanations.

Second, by self-auditing our work for potential intervention-generated inequalities, we showed that our approach, while effective on average, appears to benefit individuals with high NFC more than those with low NFC. These results add to a small but growing body of work suggesting that the user interface innovations that allow people to tackle ever more complex tasks, but which also require increasing amounts of cognitive effort to operate, systematically benefit high-NFC individuals more than those with lower levels of cognitive motivation. Thus, reducing these disparities emerges as a novel challenge for the HCI community. Our results also demonstrate the value and feasibility of auditing HCI innovations (whether in the area of explainable AI or elsewhere) for disparate effects. While there is a growing body of work on auditing the behavior of \em algorithms \em underlying interactive systems, the need to audit interaction design choices has received less attention so far but appears equally important.

\section{Related Work}

\subsection{Cognitive forcing functions and strategies}

Dual processing theory postulates that humans make decisions through one of two distinct cognitive processes --- fast and automatic (i.e., System 1 thinking) or slow and deliberative (i.e., System 2 thinking)~\cite{wason1974dual, kahneman2002representativeness, kahneman2011thinking}. System 1 thinking employs heuristics that lead to effective decisions in most daily decision-making settings. These heuristics are highly useful and mostly effective, however they can lead to systemic and predictable errors~\cite{kahneman1982judgment}. Importantly, as a definition of their expertise, experts also develop heuristics, which at times make them prone to faulty decisions even in high-stakes domains~\cite{lambe2016dual}. Therefore, shifting people's reasoning from System 1 to a more deliberative and rational, System 2 thinking remains a challenge that researchers have tackled in numerous high-stakes domains~\cite{lambe2016dual,chen2003exploratory,cooper2019cognitive}. 

For example, substantive effort has been spent in clinical settings to improve diagnostic outcomes by eliciting analytical thinking from clinicians. That is because evidence suggests that very often diagnostic errors are not cases of lack of knowledge, but rather of biases and faulty information synthesis~\cite{graber2012cognitive}. Interventions that seek to mitigate these types of errors by invoking deliberative thinking can be categorized into educational strategies and cognitive forcing functions~\cite{croskerry03:cognitive}. Educational strategies are metacognitive debiasing techniques that aim at enhancing \emph{future} decision-making by increasing clinicians awareness about the existence of different decision-making pitfalls. They are introduced through educational curricula, simulation training, and other instruction techniques. On the other hand, cognitive forcing functions are interventions which take place \emph{at the decision-making time} and encourage the decision-maker to engage analytically with the decision at hand. For instance, in clinical settings, these forcing functions can take the form of checklists, diagnostic time-outs, and slow decision-making. 
Studies indicate that cognitive forcing functions, as interventions at the time of decision-making, are more effective than educational strategies in increasing accuracy of diagnostic processes in clinical decision making~\cite{lambe2016dual,sherbino2014ineffectiveness}.

\subsection{Overreliance in AI-assisted decision-making}

As machine learning models have achieved high accuracies across numerous domains, they are increasingly being deployed as decision-support aids for humans. The implicit assumption is that because these models have high accuracies, so will the overall human+AI teams. However, mounting results suggest that these teams systematically under-perform the AI alone in tasks where AI's accuracy is higher than humans working alone (e.g., medical treatment selection~\cite{jacobs2021how}, deceptive review detection~\cite{lai2019human, lai2020why}, loan default prediction~\cite{green2019principles}, high-fat nutrient detection~\cite{bucinca20:proxy}, income category prediction~\cite{ zhang2020effect}). Many expected that if humans were shown explanations for AI decisions, the team performance would be complementary as the human would be able to understand why the AI came up with a prediction and, more importantly, detect when it was wrong. Yet recent studies found that explanations did not help in detecting AI's incorrect recommendations~\cite{jacobs2021how}. While explanations have generally improved the overall performance compared to the performance of humans alone and providing the human only with predictions, team performance is still inferior to that of AI's~\cite{bansal2020does}. A necessary next step, then, is to understand where this gain in performance (if any) comes from, and why the combined human+AI performance is still lower than the performance of AI models working independently.

A leading view to explain the inferior team performance is that humans overtrust the AI~\cite{bansal2019beyond}. In other words, when AI makes an incorrect prediction, humans rely on it even when they would have made a better decision on their own. Thus, numerous studies have highlighted the importance of \emph{calibrated trust} in improving AI-assisted decision-making~\cite{lee2004trust, jiang2018trust, bucinca20:proxy, zhang2020effect, pop2015individual, bussone2015role, parush2007degradation}. The research community is aware of the risk of overreliance on AIs in general and guidelines have been proposed to reduce overtrust. For example, Wagner et al.~\cite{wagner2018overtrust} in the context of people overtrusting robots, suggest avoiding features that may nudge the user towards the anthropomorphization of robots, or for self-driving cars, they advise developing tools that understand when the driver is not paying attention. However, to the best of our knowledge, there are no specific interventions that are designed explicitly to mitigate overreliance on AIs and that are shown empirically to reduce overtrust.

\emph{Cognitive forcing functions}, an umbrella term for interventions that elicit thinking at the decision-making time, have been implemented in various forms by prior work. Often these functions were designed to explicitly disrupt the quick, heuristic (i.e., System 1) decision-making process~\cite{chen2003exploratory,cooper2019cognitive,eberhardt2020biased,ely2011checklists,lambe2016dual,park2019slow}, but may not necessarily be referred to as \emph{cognitive forcing functions} even in the healthcare domain. Listed are some of these strategies that have either previously been employed for AI-assisted decision-making or for which we found to have appropriate analogs in the AI-assisted decision-making domain:
\begin{itemize}
    \item \emph{Asking the person to make a decision before seeing the AI's recommendation.} Prior studies in human-AI decision-making have shown the anchoring bias that occurs by presenting people with AI's recommendation before allowing them to make a decision first~\cite{green2019principles}. People made better decision when they saw the AI's recommendation after making an unassisted decision.
    \item \emph{Slowing down the process.} As shown by other HCI researchers, simply delaying the presentation of AI recommendation can improve outcomes~\cite{park2019slow}.
    \item \emph{Letting the person choose whether and when to see the AI recommendation.} There is evidence that showing unsolicited advice that contradicts a person's initial idea may trigger reactance (resistance to the advice)~\cite{fitzsimons2004reactance}. To prevent this, one could only show AI recommendations when a person requests it.
\end{itemize}

A growing number of studies demonstrate, however, that people prefer simpler constructs, even though they learn more and perform better with more complex ones. Visualization literature reveals that visual difficulties, while not necessarily preferred, improve participants' comprehension and recall of the displayed content~\cite{hullman2011benefitting}.  Recent education research also indicates that students preferred and thought they learned more with easier, passive instructions than with more cognitively demanding, active instruction. But when evaluated objectively, their actual learning and performance was better with the more cognitively demanding, active instruction~\cite{deslauriers2019measuring}. Therefore, while cognitive forcing functions may enhance user performance, there likely exists a tension with user preference of the system.


\section{Experiment}
 We conducted an experiment with 3 different cognitive forcing interventions, two simple explainable AI conditions, and a no-AI baseline, to examine whether cognitive forcing functions are successful in reducing human overreliance on the AI when working on a decision-making task. Specifically, we hypothesized that:
 
\begin{quote}
\textbf{H1a:} Compared to simple explainable AI approaches, cognitive forcing functions will improve the performance of human+AI teams in situations where the AI's top prediction is incorrect.
\end{quote}

And, consequently:

\begin{quote}
\textbf{H1b:} Compared to simple explainable AI approaches, cognitive forcing functions will improve the performance of human+AI teams.
\end{quote}
 
However, because cognitive forcing functions cause people to exert extra cognitive effort, we expected that there would be a trade-off between the acceptability of the design of the human+AI collaboration interface and its effectiveness in reducing the overreliance: people would over-rely less on the AI when they were forced to think harder, but they would prefer such interfaces less than those that require less thinking. Thus, we hypothesized that:

\begin{quote}
\textbf{H2:} There will be a negative correlation between the self-reported acceptability of the interface and the performance of human+AI teams in situations where the AI's prediction is incorrect.
\end{quote}

\subsection{Task description}

Accessing experts such as judges, or clinicians for experiments is notoriously challenging and costly. We, therefore, designed the task around nutrition as an approachable domain for laypeople. Participants were shown meal images and were asked to replace the ingredient highest in carbohydrates on the plate, with an ingredient that was low in carbohydrates, but similar in flavor. Each of the meals had an ingredient with substantially more carbohydrates than other ingredients on the plate. Participants had to make two discrete choices: First, they had to pick the ingredient to take out from the meal (by selecting from a list of approximately 10 choices, that included all the main ingredients in the image plus several other ingredients). Next, they had to select a new ingredient to put in place of the removed ingredient (again, by selecting from a list of roughly 10 choices).


\begin{figure}[t]
\begin{subfigure}{0.9\linewidth}
\includegraphics[width=\linewidth]{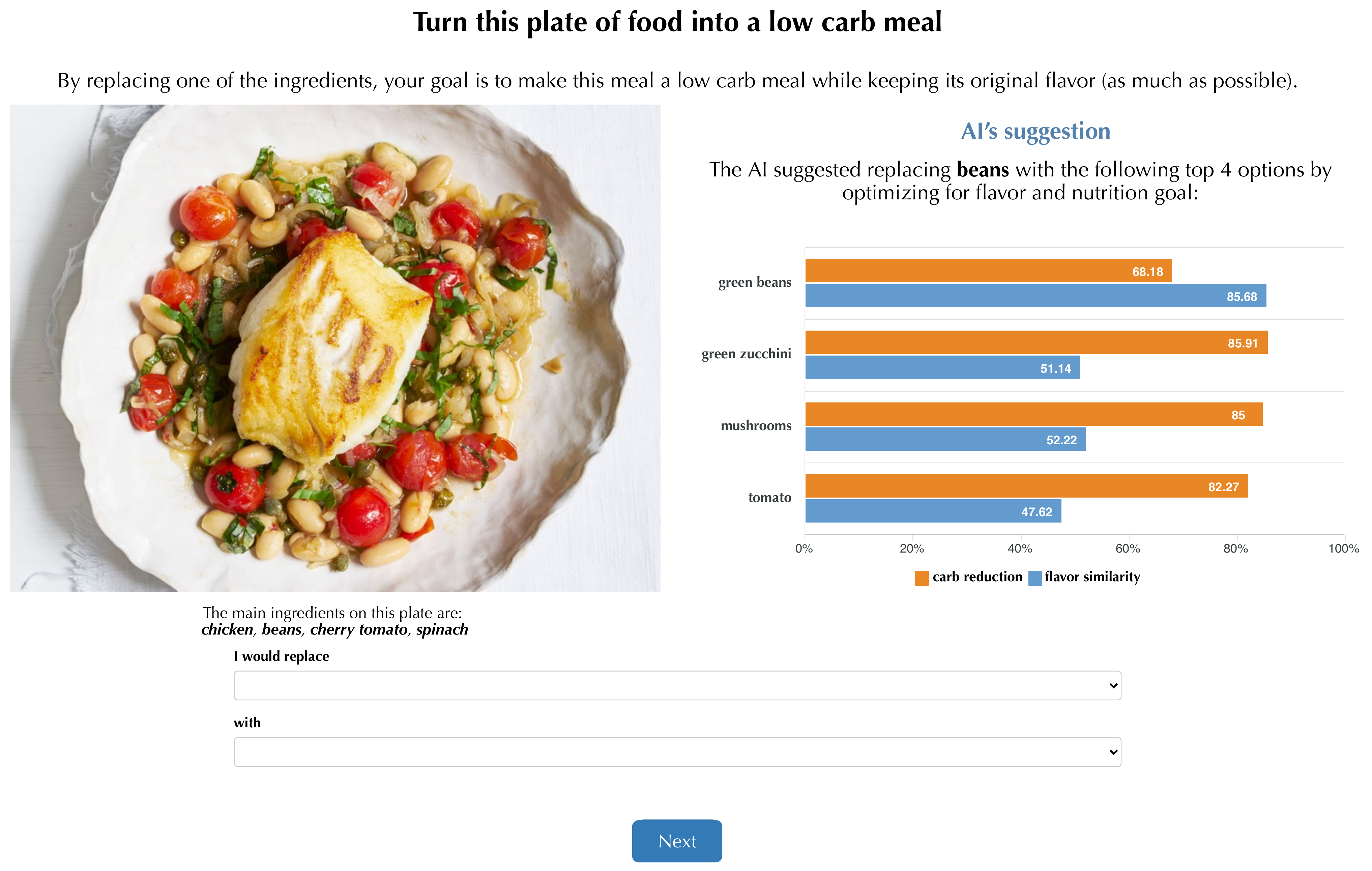}
\caption{explanation \em(SXAI)\em} \label{fig:1a}
\end{subfigure}
\hspace*{\fill} 

\begin{tabular}{lccc}
\begin{subfigure}{0.3\linewidth}
\includegraphics[width=\linewidth]{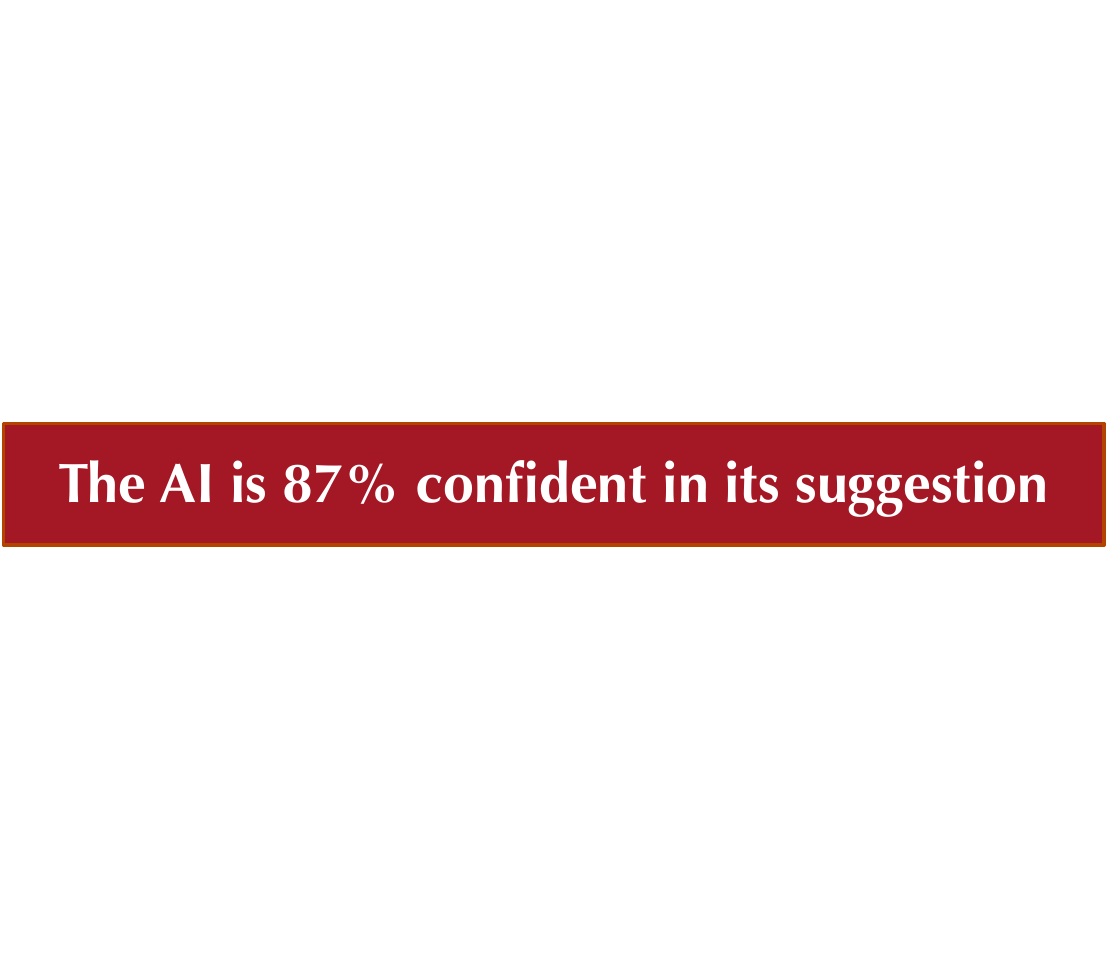}
\caption{uncertainty \em(SXAI)\em} \label{fig:1b}
\end{subfigure}
\begin{subfigure}{0.3\linewidth}
\includegraphics[width=\linewidth]{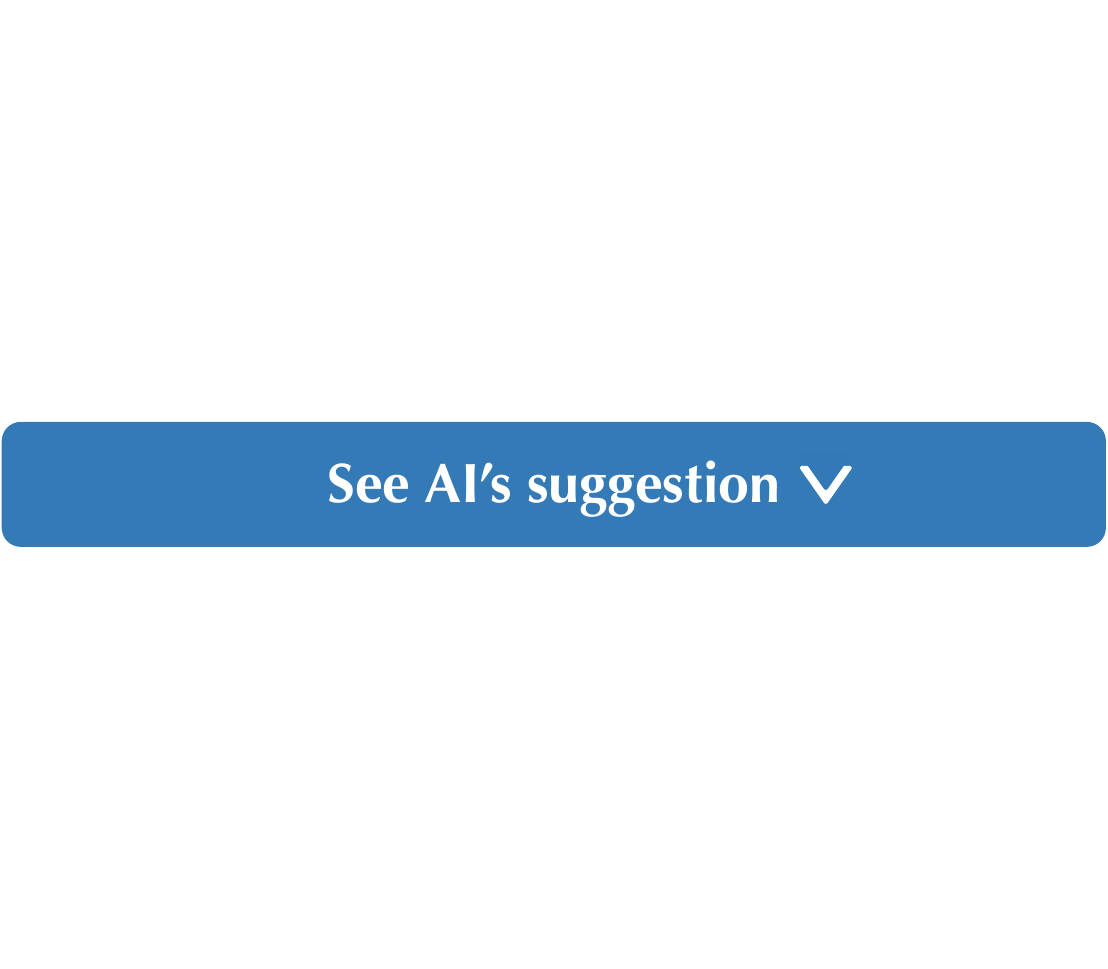}
\caption{on demand \em(CFF)\em} \label{fig:1c}
\end{subfigure}
\begin{subfigure}{0.3\linewidth}
\includegraphics[width=\linewidth]{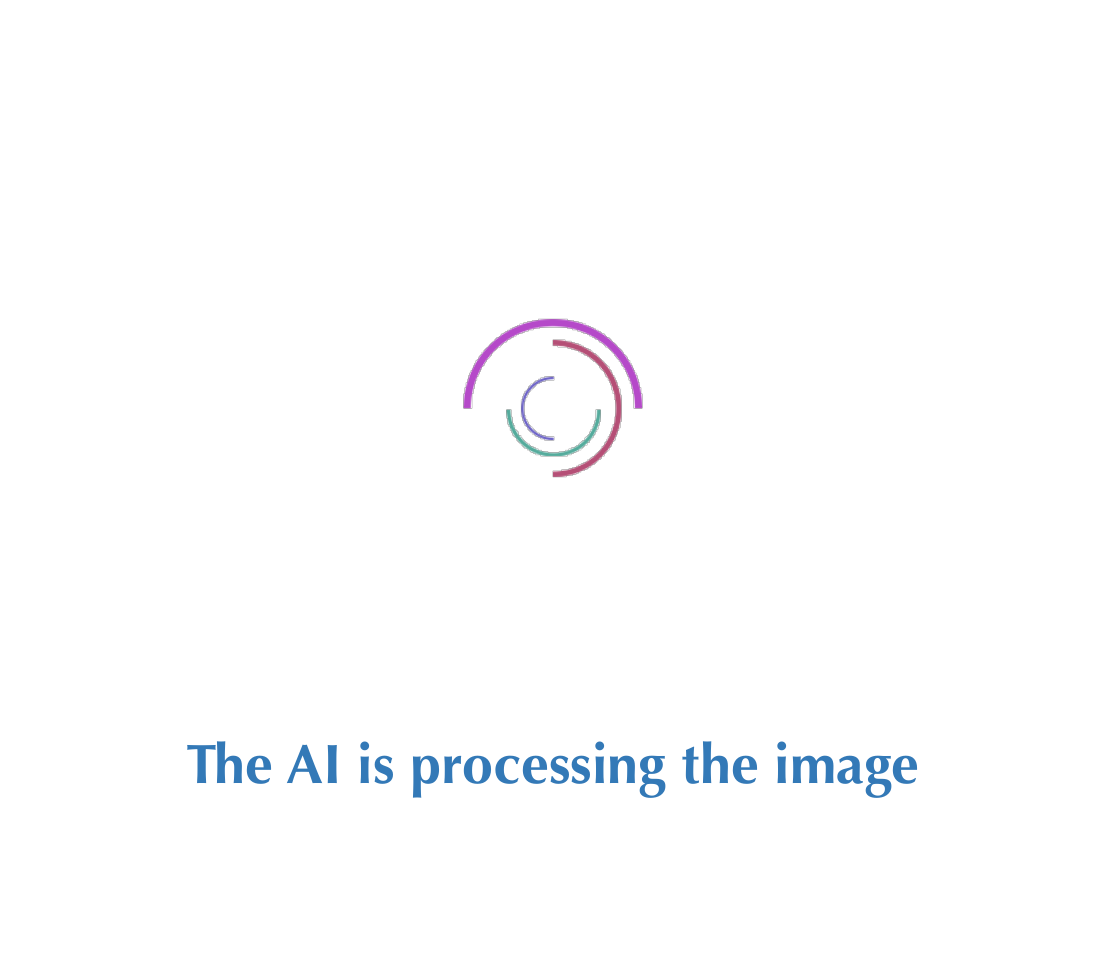}
\caption{wait \em(CFF)\em} \label{fig:1d}
\end{subfigure}
\end{tabular}
\caption{Multiple conditions. (a) depicts the main interface with the \emph{explanation} condition, where the ingredients are recognized correctly and an explanation is provided for top replacements. In \emph{uncertainty} condition (b) participants were shown AI's confidence along with the explanation. In \emph{on demand} condition (c) participants could click to see the AI's suggestion and explanation, whereas in \emph{wait} condition (d) they were shown a message ``AI is processing the image'' for 30 seconds before the suggestion and explanation were presented to them.} \label{fig:1}
\end{figure}

\subsection{Conditions}
\label{sec:conditions}
\label{conditions}
We designed six conditions that varied in whether and how a simulated AI assisted the participants in making their decisions.

\paragraph{No AI} In the \emph{no AI} condition, participants were shown just the image of the meal and the two pull-down menus to select which ingredient to remove and which new ingredient to put in its place.

\subsubsection{Simple Explainable AI (SXAI) Baselines}

\paragraph{Explanation} The \emph{explanation} condition (Figure~\ref{fig:1a}) included the list of the ingredients that the simulated AI (described in the next section) recognized on the plate and top four substitutions for the ingredient highest in carbohydrates among the recognized ingredients. Each substitution was accompanied by a specific feature-based explanation showing the estimated carbohydrate reduction and flavor similarity. While there is a great diversity in the designs of explanations in the explainable AI community, this condition reflects a common approach for designing the human-AI collaborative interface: the explanation and the AI suggestion are shown immediately to the human decision-maker.

\paragraph{Uncertainty} The \emph{uncertainty} condition was like the \emph{explanation}, except that participants were also shown a confidence prompt ``The AI is X \% confident in its suggestion.'' (Figure~\ref{fig:1b}). The prompt was shown only when the AI was uncertain about its suggestion. This condition captures another common approach for designing interfaces for human-AI collaboration~\cite{bansal2020does, lai2020why, yin2019understanding}. 

\subsubsection{Cognitive Forcing Functions (CFF)}

\paragraph{On demand} In the \emph{on demand} condition, the AI suggestion was not shown to the users by default. Users could see the suggestion and the explanation (identical to the one in the \emph{explanation} condition) if they clicked on the ``See AI's suggestion'' button (Figure~\ref{fig:1c}). We hypothesized this to be a light form of cognitive forcing function, where the human would engage with the explanation because they explicitly requested it.

\paragraph{Update} Participants in the \emph{update} condition had to make the initial decision without the help of an AI (i.e., like in the \emph{no AI} condition). Having made the initial decision, they were shown the AI's suggestion and explanation and could update their decision. We hypothesized that this condition would motivate the participants to engage with the explanation, especially when the AI disagreed with their initial decision. In other words, participants would be curious as to \emph{why} the AI disagreed with them.

\paragraph{Wait} Participants in the \emph{wait} condition were shown a message ``AI is processing the image'' (Figure~\ref{fig:1d}) for 30 seconds before being shown the AI suggestion and the explanation (identical to the one in the \emph{explanation} condition). Informed by prior work that a slow algorithm improves user's accuracy~\cite{park2019slow}, we also hypothesized this is a form of cognitive forcing function. That is, because a user forms a hypothesis for the correct answer while waiting for the AI's suggestion, then evaluates the AI explanation to check if it supports their hypothesis.

\subsection{The Simulated AI}

We designed a simulated AI for the experiment, which had 75\% accuracy of correctly recognizing the ingredient with the highest carbohydrate impact in the image of the meal. Note that we did not train an actual machine learning model for this task because we wanted to have control over the type and prevalence of error the AI would make. Once the ingredients were recognized, the simulated AI always provided suggestions for replacing the ingredient highest in carbohydrates and an explanation which included four top replacements for the ingredient, each accompanied by an estimate of carbohydrate reduction and flavor similarity. These top replacements were ranked in terms of their carb reduction and flavor similarity compared to the replaced ingredient. 

We built a lookup table for each of the ingredients present in the image and their top ten replacements. These top ten replacements were selected as the ingredients that were most similar in terms of flavor to the ingredient to be replaced. We used a flavor database comprised of flavor molecules representing an array of tastes and odors associated with 936 ingredients~\cite{garg2018flavordb}. Each ingredient is made up of different flavor molecules and the ingredients can be paired to other ingredients that share the most flavor molecules with. However, in addition to the number of flavor molecules that the ingredients share, the number of molecules that the ingredients do not share is also important to understand how similar two ingredients are. Hence, we computed the flavor similarity between two ingredients with sets of flavor molecules: $A$ and $B$ as $F_{AB} = \frac{|A \cap B|}{|A \cup B|}$. Having selected the top ten ingredients in terms of flavor, we computed the percent of carbs that would be reduced with respect to the original ingredient. We ranked the top ingredients in terms of the harmonic mean of the flavor similarity and carb reduction. We chose to compute the harmonic mean, because we sought for an optimization for both flavor and carbs, not only one of them. The AI explanations included this information about the flavor similarity and carb content of the suggested replacements. These feature based explanations (i.e., carb reduction and flavor similarity) were intended to be comparable to commonly used explanations generated by techniques such as SHAP \cite{lundberg2017unified}.
For all the conditions, once the participants selected an ingredient to replace, the second list of ingredients to replace with was populated by these top ten ingredients.

We designed errors of the simulated AI to stem from visual misrecognition. To produce incorrect model predictions, the AI would not recognize the ingredient highest in carbohydrates. Therefore, it would suggest replacing the second ingredient highest in carbs on the plate. Note that we designed the questions such that all meals included a single ingredient that had substantially more carbs than the other ingredients on the plate. As a result, the second ingredient highest in carb on the plate was a low carb ingredient.

\subsection{Procedure}
The study was conducted online on Amazon Mechanical Turk (MTurk). Participants were first presented with a consent form and instructions. They completed 26 questions, split into two blocks of 13 questions. Out of the 9 conditions (the 6 mentioned earlier, plus 3 additional exploratory designs not reported in this paper), participants were randomly shown a different condition in each block. The first question of each block was a practice question to help participants familiarize themselves with a new design, and was discarded from analysis. Six questions with incorrect model predictions (three per block) were shown overall. While the positions of these questions were fixed, the order of questions where the model predictions were correct/incorrect were randomized for each participant. Participants completed a questionnaire after each block to indicate their subjective experience with the system. After the first block, they also completed a four-item Need for Cognition questionnaire, consisting the four items with the highest factor loading from \cite{cacioppo84:efficient}.

\subsection{Participants} A total of 260 participants were recruited to complete the task via Amazon MTurk in three batches.  Participation was limited to adults residing in the United States. The study took 15 minutes on average to complete. Each participant was paid \$2.5 (USD)  for an estimated rate of \$10 per hour. Participants could take part in the study only once. To motivate participants to perform well on the task, the top performer of each batch was rewarded with a bonus of \$3. Out of 260 participants, 49 participants were filtered as they selected as an ingredient to replace one that was not on the plate for more than 55\% of the questions. We noticed that these excluded participants had also an accuracy of 0 on the task. An additional 12 participants were excluded from the analyses as they were assigned to only exploratory conditions not reported in the results.
From the retained \paricipantnumber participants, 191 completed an optional demographic survey in the beginning of the study ($M_{age}$ = 37.09, SD = 10.74). 125 of the participants (65.44\%) self-identified as male, and the rest self-identified as female. 164 of the participants (85.86\%) had either received or were pursuing a college degree, 13 of them a high school degree, and the remaining 13 were either pursuing or had received a PhD degree. For only one of the participants, the highest level of education received was a pre-high school degree.
\subsection{Design and Analysis}
The study was a mixed between- and within-subject design. Both the within subject factor and between subject factor was the condition. Each participant interacted with two of the nine conditions.

We collected the following performance measures:
\begin{itemize}
    \item Overall performance: Percentage of top replacements, including the correct ingredient to be replaced and the top replacement for it
    \item Carb source detection performance: Percentage of correct selections of ingredients to be replaced
    \item Carb reduction: Percentage of carbohydrate reduction with respect to replaced ingredients
    \item Flavor similarity: Percentage of flavor similarity with respect to replaced ingredients
    \item Overreliance: Percentage of agreement with the AI when the AI made incorrect predictions
    \item Human error (on incorrect AI predictions): Percentage of incorrect decisions (different from AI's suggestion) when the AI made incorrect predictions
\end{itemize}    

We also collected several self-reported subjective measures:
\begin{itemize}
    \item Preference: Participants rated the statement \textit{``I would like to use this system frequently.''} on a 5-point Likert scale from 1=Strongly disagree to 5=Strongly agree after each block.
    \item Trust: Participants rated the statement \textit{``I trust this AI's suggestions for optimal replacement.''} on a 5-point Likert scale from 1=Strongly disagree to 5=Strongly agree after each block. 
    \item Mental demand: Participants rated the statement \textit{``I found this task difficult.''} on a 5-point Likert scale from 1=Strongly disagree to 5=Strongly agree for each block
    \item System complexity: Participants rated the statement \textit{``The system was complex.''} on a 5-point Likert scale from 1=Strongly disagree to 5=Strongly agree for each block

\end{itemize}

We conducted our analyses first by category (\emph{no AI}, \emph{simple explainable AI (SXAI)}, and \emph{cognitive forcing functions (CFF)}). Subsequently, we conducted additional analyses to test for differences among individual designs within the SXAI and CFF categories.

For performance on incorrect model predictions, we also added \emph{no AI} category to the analysis even though participants in that category did not see any model predictions. We compared participants’ performance on questions when they saw incorrect predictions (i.e., \emph{CFF} and \emph{SXAI}) to the performance of other participants on the same questions but with no AI assistance (i.e., \emph{no AI}).

We used analysis of variance to analyze the impact of the different designs on both objective and subjective measures. Our data was analyzed using mixed-effects models. Category/condition was modeled as a fixed effect, and participant as a random effect to account for the fact that each participant saw two out of several possible conditions (i.e., all the measurements were not statistically independent) ~\cite{barr2013random}. Mixed-effects models also properly handle the imbalance in our data~\cite{spilke2005analysis}, due to participants being randomly assigned to conditions and not all participants having interacted with both categories (CFF and SXAI). Note that unbalanced data can lead to fractional denominator degrees of freedom.

We used Student's t-test for post-hoc pairwise comparisons with Holm-Bonferroni corrections to account for multiple comparisons~\cite{holm79:simple}.
We report marginal means, which are means for groups that are adjusted for means of other factors in the model (i.e., participant). The effect size (Cohen's \textit{d}) was calculated accounting for the random effect as described in Westfall et al. \cite{westfall2014statistical}.

We used Pearson's correlation to analyze associations between subjective measures and performance.

Throughout the results and discussion sections, \em significantly \em corresponds to \em statistically significantly \em for improved readability.

\section{Results}
\label{section:results}

\begin{table}[]
\resizebox{\textwidth}{!}{%
\begin{tabular}{ll|l|l|l|l|l|l}
 &
   &
  \multicolumn{1}{c|}{\textbf{Post-hoc analyses}} &
  \multicolumn{3}{c|}{\textbf{\begin{tabular}[c]{@{}c@{}}Marginal Means\\ (Standard Errors)\end{tabular}}} &
  \multicolumn{1}{c|}{\textbf{Significance}} &
  \textbf{\begin{tabular}[c]{@{}l@{}}Effect Size\\ CFF vs. SXAI\end{tabular}} \\ \cline{4-6}
 &
   &
   &
  \textbf{no AI} &
  \textbf{SXAI} &
  \textbf{CFF} &
   &
   \\ \hline
\multicolumn{1}{l|}{} &
  \textit{\textbf{all}} &
  \textbf{\{no AI\} \textless \{CFF, SXAI\}} &
  \begin{tabular}[c]{@{}l@{}}0.17\\ (0.03)\end{tabular} &
  \begin{tabular}[c]{@{}l@{}}0.35 \\ (0.02)\end{tabular} &
  \begin{tabular}[c]{@{}l@{}}0.33\\ (0.02)\end{tabular} &
  \begin{tabular}[c]{@{}l@{}}$F_{2, 141.6} = 15.43$, \\ $p \ll .0001$\end{tabular} &
  $d = .03$ \\ \cline{2-8} 
\multicolumn{1}{l|}{\multirow{-2}{*}{\textbf{\begin{tabular}[c]{@{}l@{}}Overall \\ performance\end{tabular}}}} &
  \cellcolor[HTML]{EFEFEF}\textit{\textbf{\begin{tabular}[c]{@{}l@{}}incorrect AI\\ predictions\end{tabular}}} &
  \cellcolor[HTML]{EFEFEF}\textbf{\{SXAI\} \textless \{CFF\} \textless \{no AI\}} &
  \cellcolor[HTML]{EFEFEF}\begin{tabular}[c]{@{}l@{}}0.18\\ (0.02)\end{tabular} &
  \cellcolor[HTML]{EFEFEF}\begin{tabular}[c]{@{}l@{}}0.03\\ (0.02)\end{tabular} &
  \cellcolor[HTML]{EFEFEF}\begin{tabular}[c]{@{}l@{}}0.09\\ (0.01)\end{tabular} &
  \cellcolor[HTML]{EFEFEF}\begin{tabular}[c]{@{}l@{}}$F_{2, 261.7} = 12.59$, \\ $p \ll .0001$,\end{tabular} &
  \cellcolor[HTML]{EFEFEF}\textbf{$d = .37$} \\ \hline
\multicolumn{1}{l|}{} &
  \textit{\textbf{all}} &
  \textbf{\{no AI\} \textless  \{SXAI, CFF\}} &
  \begin{tabular}[c]{@{}l@{}}0.42\\ (0.03)\end{tabular} &
  \begin{tabular}[c]{@{}l@{}}0.56\\ (0.02)\end{tabular} &
  \begin{tabular}[c]{@{}l@{}}0.57\\ (0.02)\end{tabular} &
  \begin{tabular}[c]{@{}l@{}}$F_{2, 114.4} = 12.41$, \\ $p \ll .0001$\end{tabular} &
  $d = .03$ \\ \cline{2-8} 
\multicolumn{1}{l|}{\multirow{-3}{*}{\textbf{\begin{tabular}[c]{@{}l@{}}Carb Source\\ Detection\\ Performance\end{tabular}}}} &
  \cellcolor[HTML]{EFEFEF}\textit{\textbf{\begin{tabular}[c]{@{}l@{}}incorrect AI\\ predictions\end{tabular}}} &
  \cellcolor[HTML]{EFEFEF}\textbf{\{SXAI\} \textless \{CFF\} \textless \{no AI\}} &
  \cellcolor[HTML]{EFEFEF}\begin{tabular}[c]{@{}l@{}}0.49\\ (0.04)\end{tabular} &
  \cellcolor[HTML]{EFEFEF}\begin{tabular}[c]{@{}l@{}}0.08\\ (0.03)\end{tabular} &
  \cellcolor[HTML]{EFEFEF}\begin{tabular}[c]{@{}l@{}}0.27\\ (0.03)\end{tabular} &
  \cellcolor[HTML]{EFEFEF}\begin{tabular}[c]{@{}l@{}}$F_{2, 242.6} = 35.59$, \\ $p \ll .0001$\end{tabular} &
  \cellcolor[HTML]{EFEFEF}\textbf{$d = .66$} \\ \hline
\multicolumn{1}{l|}{} &
  \textit{\textbf{all}} &
  \textbf{\{no AI\} \textless  \{SXAI, CFF\}} &
  \begin{tabular}[c]{@{}l@{}}21.73\\ (2.25)\end{tabular} &
  \begin{tabular}[c]{@{}l@{}}35.70\\ (1.81)\end{tabular} &
  \begin{tabular}[c]{@{}l@{}}35.20\\ (1.69)\end{tabular} &
  \begin{tabular}[c]{@{}l@{}}$F_{2, 123} = 20.15$, \\ $p \ll .0001$\end{tabular} &
  $d = .01$ \\ \cline{2-8} 
\multicolumn{1}{l|}{\multirow{-2}{*}{\textbf{Carb Reduction}}} &
  \cellcolor[HTML]{EFEFEF}\textit{\textbf{\begin{tabular}[c]{@{}l@{}}incorrect AI\\ predictions\end{tabular}}} &
  \cellcolor[HTML]{EFEFEF}\textbf{\{SXAI\} \textless \{CFF\} \textless \{no AI\}} &
  \cellcolor[HTML]{EFEFEF}\begin{tabular}[c]{@{}l@{}}23.58\\ (2.38)\end{tabular} &
  \cellcolor[HTML]{EFEFEF}\begin{tabular}[c]{@{}l@{}}2.96\\ (1.75)\end{tabular} &
  \cellcolor[HTML]{EFEFEF}\begin{tabular}[c]{@{}l@{}}11.37\\ (1.56)\end{tabular} &
  \cellcolor[HTML]{EFEFEF}\begin{tabular}[c]{@{}l@{}}$F_{2, 262.8} = 24.40$, \\ $p \ll .0001$\end{tabular} &
  \cellcolor[HTML]{EFEFEF}\textbf{$d = .53$} \\ \hline
\multicolumn{1}{l|}{} &
  \textit{\textbf{all}} &
  \textbf{\{no AI\} \textless  \{SXAI, CFF\}} &
  \begin{tabular}[c]{@{}l@{}}23.57\\ (2.07)\end{tabular} &
  \begin{tabular}[c]{@{}l@{}}35.26\\ (1.70)\end{tabular} &
  \begin{tabular}[c]{@{}l@{}}34.63\\ (1.60)\end{tabular} &
  \begin{tabular}[c]{@{}l@{}}$F_{2, 111.8} = 17.86$, \\ $p \ll .0001$\end{tabular} &
  $d = .01$ \\ \cline{2-8} 
\multicolumn{1}{l|}{\multirow{-2}{*}{\textbf{Flavor Similarity}}} &
  \cellcolor[HTML]{EFEFEF}\textit{\textbf{\begin{tabular}[c]{@{}l@{}}incorrect AI\\ predictions\end{tabular}}} &
  \cellcolor[HTML]{EFEFEF}\textbf{\{SXAI\} \textless \{CFF\} \textless \{no AI\}} &
  \cellcolor[HTML]{EFEFEF}\begin{tabular}[c]{@{}l@{}}24.00\\ (2.23)\end{tabular} &
  \cellcolor[HTML]{EFEFEF}\begin{tabular}[c]{@{}l@{}}3.37\\ (1.64)\end{tabular} &
  \cellcolor[HTML]{EFEFEF}\begin{tabular}[c]{@{}l@{}}11.28\\ (1.46)\end{tabular} &
  \cellcolor[HTML]{EFEFEF}\begin{tabular}[c]{@{}l@{}}$F_{2, 262.1} = 27.95$, \\ $p \ll .0001$\end{tabular} &
  \cellcolor[HTML]{EFEFEF}\textbf{$d = .53$} \\ \hline
\end{tabular}
}
\caption{Results of objective measures. Differences indicated by the \textless\ symbol are statistically significant. Categories that share brackets do not differ significantly.} \label{table:objective measures}
\end{table}

\begin{table}[]
\resizebox{\textwidth}{!}{%
\begin{tabular}{lll|l|l|l}
 &
   &
  \multicolumn{1}{c|}{\textbf{}} &
  \multicolumn{2}{c|}{\textbf{\begin{tabular}[c]{@{}c@{}}Marginal Means\\ (Standard Errors)\end{tabular}}} &
  \multicolumn{1}{c}{\textbf{\begin{tabular}[c]{@{}c@{}}Significance\\ \& Effect Size\end{tabular}}} \\ \cline{4-5}
 &
   &
   &
  \textbf{SXAI} &
  \textbf{CFF} &
   \\ \Xhline{2\arrayrulewidth}
\multicolumn{1}{l|}{} &
  \multicolumn{1}{l|}{\textit{\textbf{correct}}} &
  \textbf{\{SXAI\} \textless \{CFF\}} &
  \begin{tabular}[c]{@{}l@{}}0.03\\ (0.02)\end{tabular} &
  \begin{tabular}[c]{@{}l@{}}0.09\\ (0.01)\end{tabular} &
  \begin{tabular}[c]{@{}l@{}}$F_{1, 207.6} = 8.95$, $p = .003$, \\ $d = .37$\end{tabular} \\ \cline{2-6} 

\multicolumn{1}{l|}{} &
  \multicolumn{1}{l|}{\textit{\textbf{overrelied}}} &
  \{CFF, SXAI\} &
  \begin{tabular}[c]{@{}l@{}}0.30\\ (0.03)\end{tabular} &
  \begin{tabular}[c]{@{}l@{}}0.26\\ (0.03)\end{tabular} &
  \begin{tabular}[c]{@{}l@{}}$F_{1, 180.9} = 1.02$, $n.s.$, \\ $d = .12$\end{tabular} \\ \cline{2-6} 
\multicolumn{1}{l|}{\multirow{-3}{*}{\textbf{Overall}}} &
  \multicolumn{1}{l|}{\textit{\textbf{human error}}} &
  \{CFF, SXAI\} &
  \begin{tabular}[c]{@{}l@{}}0.68\\ (0.03)\end{tabular} &
  \begin{tabular}[c]{@{}l@{}}0.65\\ (0.03)\end{tabular} &
  \begin{tabular}[c]{@{}l@{}}$F_{1, 95.5} = 0.12$, $n.s.$, \\ $d = .07$\end{tabular} \\ \Xhline{2\arrayrulewidth}
\multicolumn{1}{l|}{} &
  \multicolumn{1}{l|}{\textit{\textbf{correct}}} &
  \textbf{\{SXAI\} \textless \{CFF\}} &
  \begin{tabular}[c]{@{}l@{}}0.08\\ (0.03)\end{tabular} &
 \begin{tabular}[c]{@{}l@{}}0.27\\ (0.03)\end{tabular} &
  \begin{tabular}[c]{@{}l@{}}$F_{1, 197.4} = 24.11$, $p \ll .0001$, \\ $d = .66$\end{tabular} \\ \cline{2-6} 
\multicolumn{1}{l|}{} &
  \multicolumn{1}{l|}{\textit{\textbf{overrelied}}} &
 \textbf{\{CFF\} \textless \{SXAI\}} &
  \begin{tabular}[c]{@{}l@{}}0.64\\ (0.03)\end{tabular} &
  \begin{tabular}[c]{@{}l@{}}0.48\\ (0.03)\end{tabular} &
 \begin{tabular}[c]{@{}l@{}}$F_{1, 145.8} = 9.24$, $p = .003$, \\ $d = .36$\end{tabular} \\ \cline{2-6}
\multicolumn{1}{l|}{\multirow{-3}{*}{\textbf{\begin{tabular}[c]{@{}l@{}}Carb Source \\ Detection\end{tabular}}}} &
  \multicolumn{1}{l|}{\textit{\textbf{human error}}} &
  \{CFF, SXAI\} &
  \begin{tabular}[c]{@{}l@{}}0.27\\ (0.04)\end{tabular} &
  \begin{tabular}[c]{@{}l@{}}0.26\\ (0.03)\end{tabular} &
  \begin{tabular}[c]{@{}l@{}}$F_{1, 95.5} = 0.12$, $n.s.$, \\ $d = .04$\end{tabular} \\ \Xhline{2\arrayrulewidth}
\end{tabular}}
\caption{Distribution of decisions on instances where the model was incorrect. Differences indicated by the \textless\ symbol are statistically significant. Categories that share brackets do not differ significantly.}
\label{overreliance}
\end{table}

\subsection{Objective measures}

Results for the performance measures are summarized in Table~\ref{table:objective measures}. Participants had to pick both the ingredient to replace and an ingredient to replace it with, each from a list that contained an average of 10 ingredients. Therefore, the probability that a participant was correct by chance in this task was approximately 1\% for the overall decision, and 10\% for carb source detection. Note that we use the term \em correct \em when referring to the \em optimal \em ingredient replacement in terms of flavor similarity and carb reduction for the ingredient highest in carbs on the plate.

When analyzing performance on all task instances (both those where the top AI predictions were correct and those where they were incorrect), both \emph{simple explainable AI} conditions and \emph{cognitive forcing functions} improved participants' performance on all measures (overall performance, carb source detection, carb reduction, and flavor similarity) compared to the \emph{no AI} baseline. There were no significant differences  on any of these metrics between \emph{cognitive forcing functions} and \emph{simple explainable AI}. There were also no significant differences within categories.

When the top AI predictions were incorrect, however, \emph{cognitive forcing functions} improved the objective metrics (i.e., overall performance, carb source detection performance, carb reduction and flavor similarity) significantly more compared to \emph{simple explainable AI}. Yet, performance of participants that completed the task with no AI assistance---\emph{no AI} category---was significantly higher than that of participants' in either \emph{cognitive forcing functions} or \emph{simple explainable AI} categories when they saw incorrect model predictions. There were no significant differences within categories.

\paragraph{\textbf{Detailed analysis of decisions when model predictions were incorrect.}}
On incorrect model predictions, participants could either follow the incorrect AI suggestion (i.e., overrely), provide a different incorrect answer (i.e., human error), or provide a correct answer. 

Distributions of overreliance, human error, and correctness were significantly different across categories for overall performance ($\chi^2(2, N=663) = 13.30$, $p = .0013$). As shown in Table~\ref{overreliance}, participants in \emph{cognitive forcing functions} made significantly more correct decisions than participants in \emph{simple explainable AI}. They also overrelied less, but not significantly so. There were also no significant differences between categories for human errors. 

For carb source detection,  distributions of overreliance, human error, and correctness were also significantly different across categories ($\chi^2(2, N=663) = 44.35$, $p \ll .0001$). As shown in Table~\ref{overreliance}, participants in \emph{cognitive forcing functions} overrelied significantly less and made significantly more correct decisions than participants in \emph{simple explainable AI}. There were no significant differences between categories for human errors. 

For all the metrics, there were no significant differences among conditions within either category (i.e., \emph{simple explainable AI} and \emph{cognitive forcing functions}).

\begin{table}[]
\resizebox{\textwidth}{!}{
\begin{tabular}{l|l|l|l|l|l}
 &
  \multicolumn{1}{c|}{} &
  \multicolumn{3}{c|}{\textbf{\begin{tabular}[c]{@{}c@{}}Marginal Means\\ (Standard Errors)\end{tabular}}} &
  \multicolumn{1}{c}{\textbf{Significance}} \\ \cline{3-5}
 &
  \multicolumn{1}{c|}{\multirow{-2}{*}{\textbf{Post-hoc analyses}}} &
  \textbf{no AI} &
  \textbf{SXAI} &
  \textbf{CFF} &
  \textbf{} \\ \hline
\textbf{Trust} &
  \{CFF, SXAI\} &
  \multicolumn{1}{c|}{/} &
  \begin{tabular}[c]{@{}l@{}}3.91\\ (0.09)\end{tabular} &
  \begin{tabular}[c]{@{}l@{}}3.72\\ (0.08)\end{tabular} &
  $F_{1, 146.4} = 2.84$, $n.s.$ \\ \hline
\textbf{Preference} &
  \textbf{\{no AI\} \textless \{CFF, SXAI\}} &
  \begin{tabular}[c]{@{}l@{}}3.13\\ (0.12)\end{tabular} &
  \begin{tabular}[c]{@{}l@{}}3.78\\ (0.09)\end{tabular} &
  \begin{tabular}[c]{@{}l@{}}3.62\\ (0.09)\end{tabular} &
  $F_{2, 169.5} = 10.52$, $p \ll .0001$ \\ \hline
\textbf{Mental Demand} &
  \textbf{\{SXAI, CFF\} \textless \{no AI\}} &
  \begin{tabular}[c]{@{}l@{}}3.66\\ (0.16)\end{tabular} &
  \begin{tabular}[c]{@{}l@{}}2.86\\ (0.12)\end{tabular} &
  \begin{tabular}[c]{@{}l@{}}2.84\\ (0.11)\end{tabular} &
  $F_{2, 183.4} = 11.70$, $p \ll .0001$ \\ \hline
\textbf{System Complexity} &
  \textbf{\{SXAI\} \textless \{CFF, no AI\}} &
  \begin{tabular}[c]{@{}l@{}}3.19\\ (0.14)\end{tabular} &
  \begin{tabular}[c]{@{}l@{}}2.64\\ (0.11)\end{tabular} &
  \begin{tabular}[c]{@{}l@{}}2.95\\ (0.10)\end{tabular} &
  $F_{2, 172.2} = 6.19$, $p = .002$
\end{tabular}
}
\caption{Results of subjective measures. Differences indicated by the \textless\ symbol are statistically significant. Categories that share brackets do not differ significantly.}\label{table:subjective measures}
\end{table}

\subsection{Subjective measures}

Table \ref{table:subjective measures} summarizes the comparisons of subjective measures across categories. 

Overall, participants reported higher trust in the AI in \emph{simple explainable AI} conditions compared to \emph{cognitive forcing functions}, albeit not significantly so.
They preferred completing the task significantly more with AI assistance (either with \emph{simple explainable AI} or \emph{cognitive forcing functions}), than without it (\emph{no AI}). They found the \emph{no AI} condition to be significantly more mentally demanding than \emph{cognitive forcing functions} and \emph{simple explainable AI} conditions. They perceived the system as significantly less complex in \emph{simple explainable AI} conditions compared to either \emph{cognitive forcing functions} or \emph{no AI}. 
Conditions within categories did not differ significantly on any of the subjective ratings.

\subsection{Subjective measures vs. objective measures}
\begin{figure}
\begin{subfigure}{\linewidth}
\includegraphics[width=\linewidth]{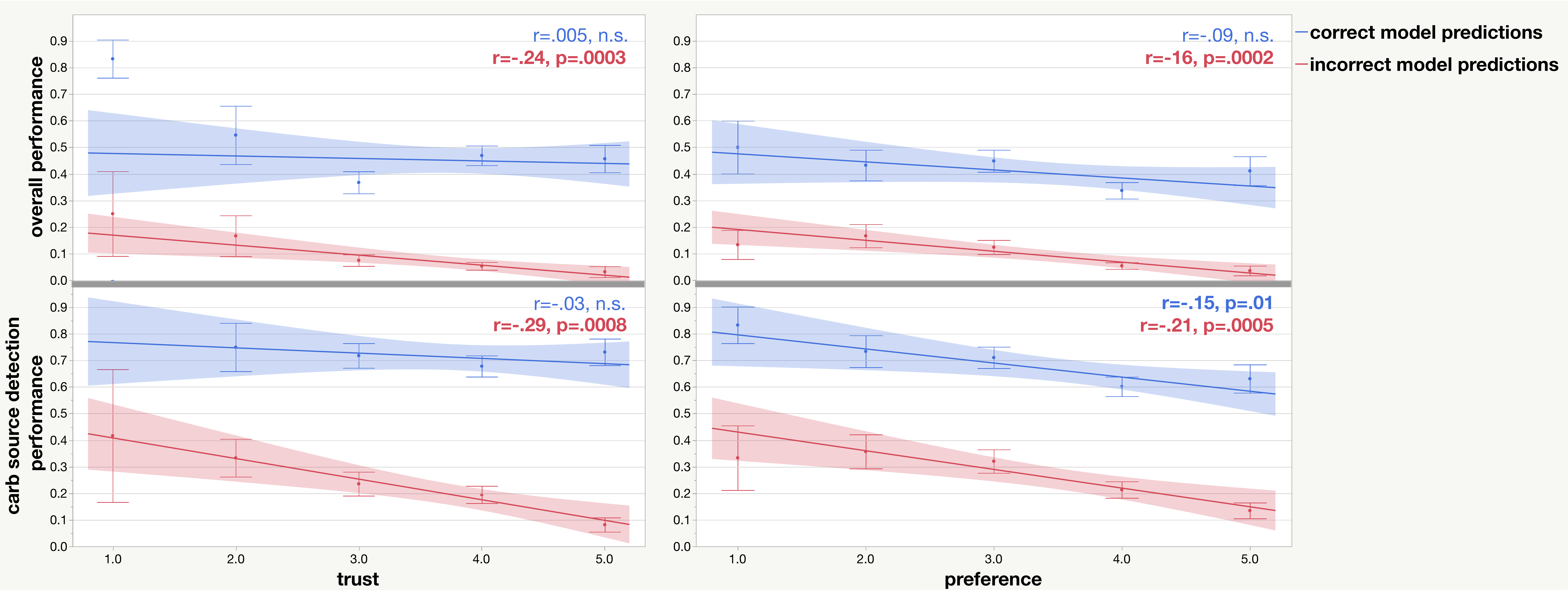}
\caption{Trust vs. performance (left) and preference vs. performance (right)} \label{fig:trust vs. performance} 
\end{subfigure}

\begin{subfigure}{\linewidth}
\includegraphics[width=\linewidth]{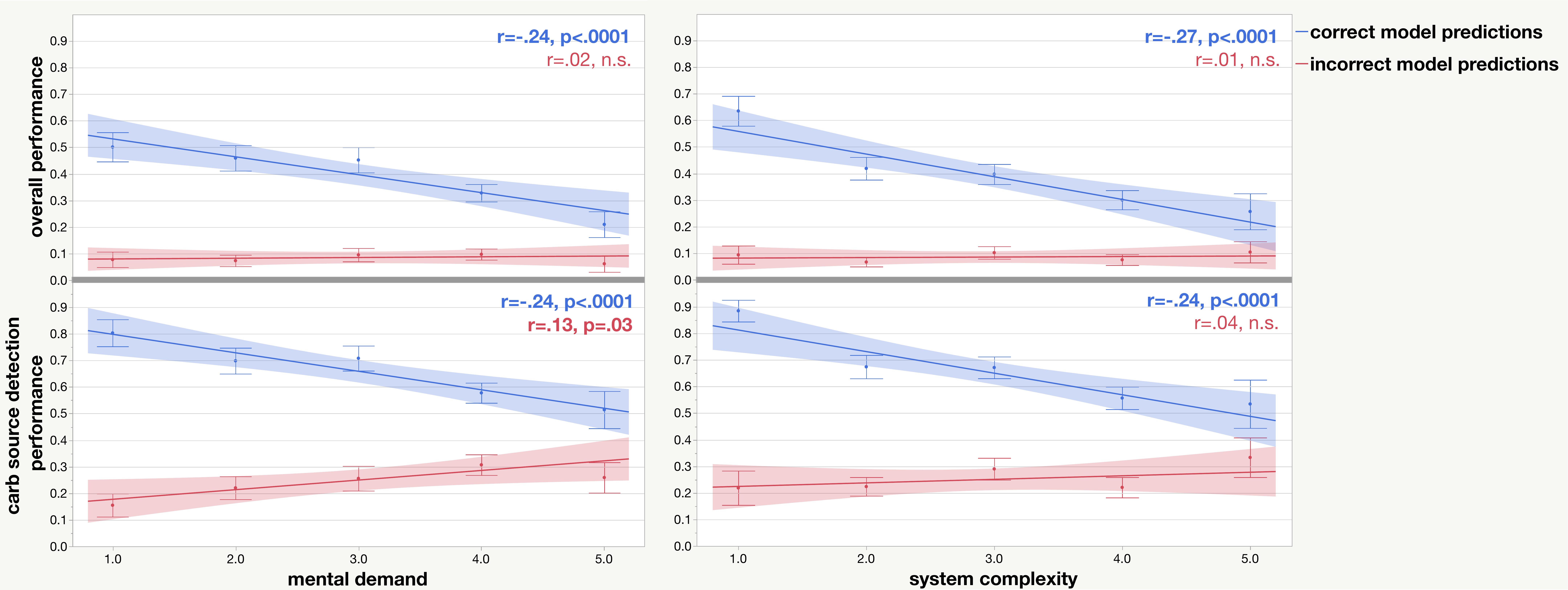}
\caption{Mental demand vs. performance (left) and system complexity vs. performance (right)} \label{fig:mental demand vs. performance} 
\end{subfigure}

\caption{Subjective measures vs. performance}
\label{performance-vs-subjective}
\end{figure}

Figure \ref{performance-vs-subjective} depicts relationships between subjective measures and performance. 

Trust and preference were significantly negatively correlated with both overall performance and carb source detection performance for incorrect model predictions. Preference was also significantly negatively correlated with carb source detection for correct model predictions.

Mental demand and system complexity were significantly negatively correlated with overall and carb source detection performance for correct model predictions. However, mental demand was significantly \em positively \em correlated with carb source detection performance for incorrect model predictions. 

Trust was significantly positively correlated with reliance on incorrect model predictions (i.e., overreliance) for carb source detection, but not so for correct model predictions. Trust was not significantly correlated with reliance for overall decision either for correct or incorrect model predictions.


\begin{figure}
\includegraphics[width=0.5\linewidth]{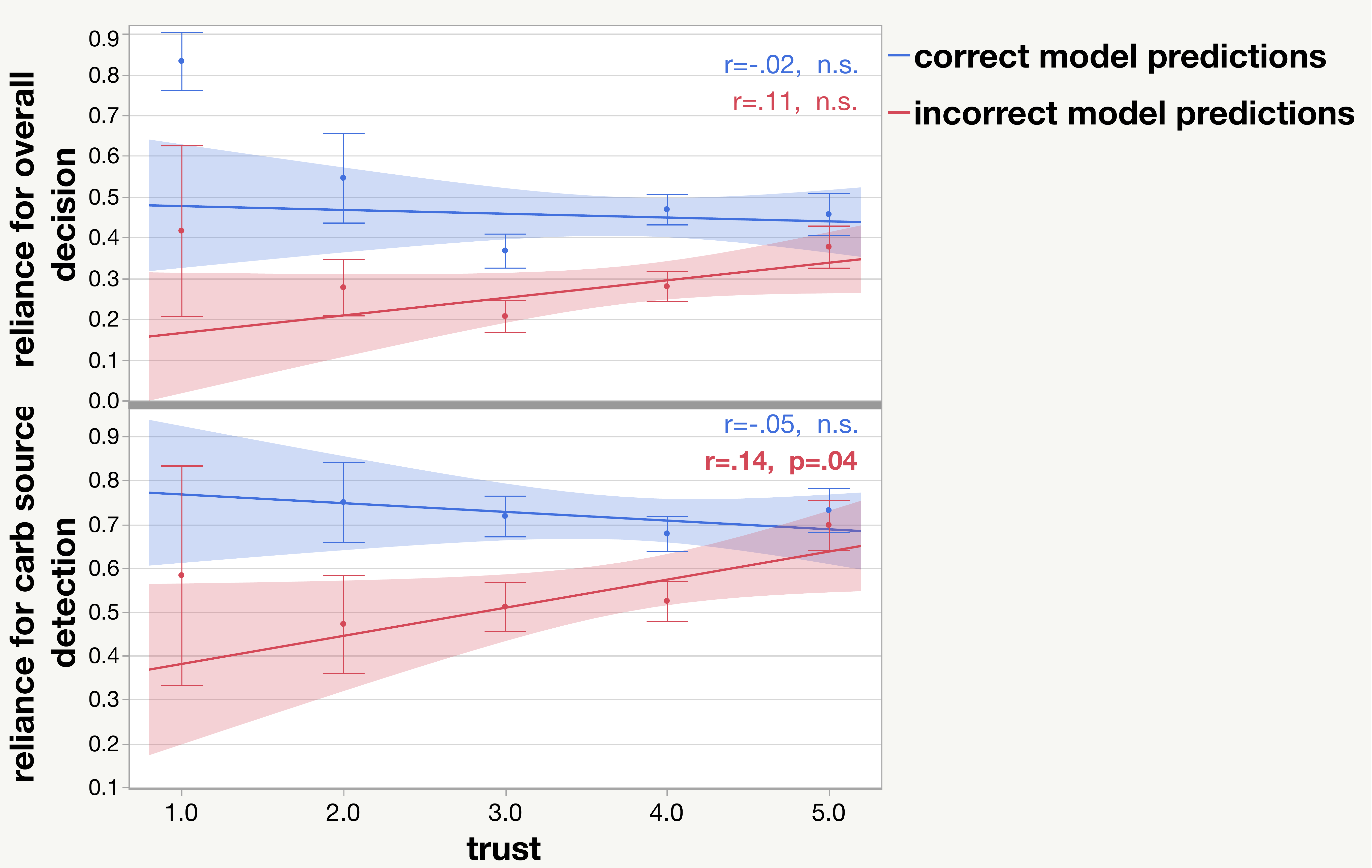}
\caption{Trust vs. reliance on AI for correct model predictions and incorrect model predictions (i.e., overreliance)}
\label{trust-vs-reliance}
\end{figure}

\section{Ethical considerations: Individual differences in cognitive motivation}

We are mindful of the fact that technological interventions can lead to intervention-generated inequalities (IGIs). IGIs arise when the benefits of an intervention disproportionately accrue to a group that is already privileged in a particular context~\cite{veinot2018good}. Thus, even though everyone might benefit to some extent, the gaps between different groups increase.
 
Therefore, we conducted a limited internal audit~\cite{raji2020closing} of our work to check for inequalities that might emerge and to understand how different groups of people are affected by the introduction of cognitive forcing functions into the AI-assisted decision-making processes. 

While some prior research in ethical AI documented disparities by disaggregating results by race and gender~\cite{buolamwini2018gender}, we believe that the relevant variable in our case is the intrinsic cognitive motivation. In psychology, it is captured through the concept of Need for Cognition (NFC), a stable personality trait that reflects how much a person enjoys engaging in cognitively demanding activities~\cite{cacioppo82:need,petty86:elaboration}. 

The impact of NFC on cognitive engagement with information has been studied in fields such as advertising and purchasing behavior~\cite{haugtvedt92:need,lin11:effects,lassiter91:need}, skill acquisition~\cite{cazan14:need}, web usage~\cite{tuten01:social,sicilia05:effects}, team function~\cite{kearney09:when}, health communication~\cite{vidrine07:construction,williams-piehota03:matching}, and AI-assisted decision-making~\cite{ghai2020explainable}. The evidence that has accrued is very consistent: high-NFC participants seek out more information and process it more deeply, while low-NFC participants are more likely to resort to cognitive shortcuts such as relying on the surface cues to assess the information such as the authority or celebrity of the source of the information, or the aesthetics of the presentation. In the context of HCI, people with high NFC are more likely to adopt novel productivity enhancing features in complex software~\cite{carenini01:analysis}. Also, given multiple ways of getting a task done, high-NFC participants are more likely than those with low NFC to choose the method that saves manual effort, but requires increased cognitive exertion~\cite{gajos17:influence}. However, there is also some evidence that people with high NFC benefit less from explanations (in terms of confidence in their decisions) when using recommender systems than people with low NFC~\cite{millecamp2019explain} although there appear to be limits to the generalizability of this result~\cite{millecamp2020whats}.

Therefore, in the context of human-AI collaboration on decision making we considered individuals with high NFC to be the already privileged group and we investigated whether \emph{cognitive forcing functions} were equally effective for people with low NFC as they were for people with high NFC or whether they increased the performance gap (i.e., inequality) between the two groups. In anticipation of this analysis, we had included a 4-item subset of the NFC questionnaire~\cite{cacioppo84:efficient} as part of our study (we used the same subset as~\cite{gajos17:influence}).

\begin{table}[]
\resizebox{\textwidth}{!}{
\begin{tabular}{ll|l|l|l|l|l|l|l|l}
 &
   &
  \multicolumn{4}{c|}{\textbf{High NFC}} &
  \multicolumn{4}{c}{\textbf{Low NFC}} \\ \cline{3-10} 
 &
   &
  \multicolumn{1}{c|}{\textbf{}} &
  \multicolumn{2}{c|}{\textbf{\begin{tabular}[c]{@{}c@{}}Marginal Means\\ (Standard Errors)\end{tabular}}} &
  \multicolumn{1}{c|}{\textbf{\begin{tabular}[c]{@{}c@{}}Significance\\ \& Effect Size\end{tabular}}} &
  \multicolumn{1}{c|}{\textbf{}} &
  \multicolumn{2}{c|}{\textbf{\begin{tabular}[c]{@{}c@{}}Marginal Means\\ (Standard Errors)\end{tabular}}} &
  \multicolumn{1}{c}{\textbf{\begin{tabular}[c]{@{}c@{}}Significance\\ \& Effect Size\end{tabular}}} \\ \cline{4-5} \cline{8-9}
 &
   &
   &
  \textbf{SXAI} &
  \textbf{CFF} &
   &
   &
  \textbf{SXAI} &
  \textbf{CFF} &
   \\ \hline
\multicolumn{1}{l|}{} &
  \textit{\textbf{all}} &
  \{CFF, SXAI\} &
  \begin{tabular}[c]{@{}l@{}}0.44\\ (0.03)\end{tabular} &
  \begin{tabular}[c]{@{}l@{}}0.43\\ (0.03)\end{tabular} &
  \begin{tabular}[c]{@{}l@{}}$F_{1, 40.65} = 0.03$, \\ $n.s.$, $d = .01$\end{tabular} &
  \{CFF, SXAI\} &
  \begin{tabular}[c]{@{}l@{}}0.25\\ (0.04)\end{tabular} &
  \begin{tabular}[c]{@{}l@{}}0.21\\ (0.03)\end{tabular} &
  \begin{tabular}[c]{@{}l@{}}$F_{1, 57.42} = 1.06$, \\ $n.s.$, $d = .16$\end{tabular} \\ \cline{2-10} 
\multicolumn{1}{l|}{\multirow{-2}{*}{\textbf{\begin{tabular}[c]{@{}l@{}}Overall \\ performance\end{tabular}}}} &
  \cellcolor[HTML]{EFEFEF}\textit{\textbf{\begin{tabular}[c]{@{}l@{}}incorrect AI\\ predictions\end{tabular}}} &
  \cellcolor[HTML]{EFEFEF}\textbf{\{SXAI\} \textless\ \{CFF\}} &
  \cellcolor[HTML]{EFEFEF}\begin{tabular}[c]{@{}l@{}}0.02\\ (0.02)\end{tabular} &
  \cellcolor[HTML]{EFEFEF}\begin{tabular}[c]{@{}l@{}}0.11\\ (0.02)\end{tabular} &
  \cellcolor[HTML]{EFEFEF}\begin{tabular}[c]{@{}l@{}}$F_{1, 100.6} = 10.55$,\\ $p = .002$, $d = .57$\end{tabular} &
  \cellcolor[HTML]{EFEFEF}\{SXAI, CFF\} &
  \cellcolor[HTML]{EFEFEF}\begin{tabular}[c]{@{}l@{}}0.03\\ (0.02)\end{tabular} &
  \cellcolor[HTML]{EFEFEF}\begin{tabular}[c]{@{}l@{}}0.06\\ (0.02)\end{tabular} &
  \cellcolor[HTML]{EFEFEF}\begin{tabular}[c]{@{}l@{}}$F_{1, 99.37} = 0.94$,\\ $n.s.$, $d = .15$\end{tabular} \\ \hline
\multicolumn{1}{l|}{} &
  \textit{\textbf{all}} &
  \{SXAI, CFF\} &
  \begin{tabular}[c]{@{}l@{}}0.66\\ (0.03)\end{tabular} &
  \begin{tabular}[c]{@{}l@{}}0.68\\ (0.03)\end{tabular} &
  \begin{tabular}[c]{@{}l@{}}$F_{1, 48} = 0.58$, \\ $n.s.$, $d = .09$\end{tabular} &
  \{CFF, SXAI\} &
  \begin{tabular}[c]{@{}l@{}}0.45\\ (0.04)\end{tabular} &
  \begin{tabular}[c]{@{}l@{}}0.44\\ (0.04)\end{tabular} &
  \begin{tabular}[c]{@{}l@{}}$F_{1, 36.95} = 0.12$, \\ $n.s.$, $d = .16$\end{tabular} \\ \cline{2-10} 
\multicolumn{1}{l|}{\multirow{-3}{*}{\textbf{\begin{tabular}[c]{@{}l@{}}Carb Source\\ Detection\\ Performance \end{tabular}}}} &
  \cellcolor[HTML]{EFEFEF}\textit{\textbf{\begin{tabular}[c]{@{}l@{}}incorrect AI\\ predictions\end{tabular}}} &
  \cellcolor[HTML]{EFEFEF}\textbf{\{SXAI\} \textless\ \{CFF\}} &
  \cellcolor[HTML]{EFEFEF}\begin{tabular}[c]{@{}l@{}}0.08\\ (0.04)\end{tabular} &
  \cellcolor[HTML]{EFEFEF}\begin{tabular}[c]{@{}l@{}}0.29\\ (0.04)\end{tabular} &
  \cellcolor[HTML]{EFEFEF}\begin{tabular}[c]{@{}l@{}}$F_{1, 90.51} = 14.75$, \\ $p = .0002$, $d = .69$\end{tabular} &
  \cellcolor[HTML]{EFEFEF}\textbf{\{SXAI\} \textless\ \{CFF\}} &
  \cellcolor[HTML]{EFEFEF}\begin{tabular}[c]{@{}l@{}}0.09\\ (0.04)\end{tabular} &
  \cellcolor[HTML]{EFEFEF}\begin{tabular}[c]{@{}l@{}}0.26\\ (0.04)\end{tabular} &
  \cellcolor[HTML]{EFEFEF}\begin{tabular}[c]{@{}l@{}}$F_{1, 99.15} = 9.39$,\\ $p = .003$, $d = .57$\end{tabular} \\ \hline
\end{tabular}
}
\caption{Performance disaggregated by NFC. Differences indicated by the \textless\ symbol are statistically significant. Categories that share brackets do not differ significantly.}
\label{table:performance-NFC}
\end{table}

\begin{table}[]
\resizebox{\textwidth}{!}{
\begin{tabular}{ll|l|l|l|l|l|l|l|l}
 &
  \multicolumn{1}{c|}{\textbf{}} &
  \multicolumn{4}{c|}{\textbf{High NFC}} &
  \multicolumn{4}{c}{\textbf{Low NFC}} \\ \cline{3-10} 
 &
   &
  \multicolumn{1}{c|}{\textbf{}} &
  \multicolumn{2}{c|}{\textbf{\begin{tabular}[c]{@{}c@{}}Marginal Means\\ (Standard Errors)\end{tabular}}} &
  \multicolumn{1}{c|}{} &
  \multicolumn{1}{c|}{} &
  \multicolumn{2}{c|}{\textbf{\begin{tabular}[c]{@{}c@{}}Marginal Means\\ (Standard Errors)\end{tabular}}} &
  \multicolumn{1}{c}{} \\ \cline{4-5} \cline{8-9}
 &
   &
   &
  \textbf{SXAI} &
  \textbf{CFF} &
  \multicolumn{1}{c|}{\multirow{-2}{*}{\textbf{\begin{tabular}[c]{@{}c@{}}Significance\\ \& Effect Size\end{tabular}}}} &
  \multicolumn{1}{c|}{\multirow{-2}{*}{\textbf{}}} &
  \textbf{SXAI} &
  \textbf{CFF} &
  \multicolumn{1}{c}{\multirow{-2}{*}{\textbf{\begin{tabular}[c]{@{}c@{}}Significance\\ \& Effect Size\end{tabular}}}} \\ \Xhline{2\arrayrulewidth}
\multicolumn{1}{l|}{} &
  \textit{\textbf{correct}} &
  \textbf{\{SXAI\} \textless \{CFF\}} &
  \begin{tabular}[c]{@{}l@{}}0.02\\ (0.02)\end{tabular} &
  \begin{tabular}[c]{@{}l@{}}0.11\\ (0.02)\end{tabular} &
  \begin{tabular}[c]{@{}l@{}}$F_{1, 100.6} = 10.55$,\\ $p = .002$, $d = .57$\end{tabular} &
  \{SXAI, CFF\} &
  \begin{tabular}[c]{@{}l@{}}0.03\\ (0.02)\end{tabular} &
  \begin{tabular}[c]{@{}l@{}}0.07\\ (0.02)\end{tabular} &
  \begin{tabular}[c]{@{}l@{}}$F_{1, 99.37} = 0.94$, \\ $n.s.$, $d = .15$\end{tabular} \\ \cline{2-10} 
\multicolumn{1}{l|}{} &
  \textit{\textbf{overrelied}} &
  \{CFF, SXAI\} &
  \begin{tabular}[c]{@{}l@{}}0.39\\ (0.05)\end{tabular} &
  \begin{tabular}[c]{@{}l@{}}0.32\\ (0.04)\end{tabular} &
  \begin{tabular}[c]{@{}l@{}}$F_{1, 82.55} = 1.13$, \\ $n.s.$, $d = .18$\end{tabular} &
  \{CFF, SXAI\} &
  \begin{tabular}[c]{@{}l@{}}0.22\\ (0.04)\end{tabular} &
  \begin{tabular}[c]{@{}l@{}}0.19\\ (0.04)\end{tabular} &
  \begin{tabular}[c]{@{}l@{}}$F_{1, 96.76} = 0.26$, \\ $n.s.$, $d = .09$\end{tabular} \\ \cline{2-10} 
\multicolumn{1}{l|}{\multirow{-3}{*}{\textbf{\begin{tabular}[c]{@{}l@{}}Overall \\ performance\end{tabular}}}} &
  \textit{\textbf{human error}} &
  \{CFF, SXAI\} &
  \begin{tabular}[c]{@{}l@{}}0.60\\ (0.05)\end{tabular} &
  \begin{tabular}[c]{@{}l@{}}0.57\\ (0.04)\end{tabular} &
  \begin{tabular}[c]{@{}l@{}}$F_{1, 75.04} = 0.35$,\\  $n.s.$, $d = .09$\end{tabular} &
  \{CFF, SXAI\} &
  \begin{tabular}[c]{@{}l@{}}0.75\\ (0.05)\end{tabular} &
  \begin{tabular}[c]{@{}l@{}}0.77\\ (0.05)\end{tabular} &
  \begin{tabular}[c]{@{}l@{}}$F_{1, 69.76} = 0.07$, \\ $n.s.$, $d = .05$\end{tabular} \\ \Xhline{2\arrayrulewidth}
\multicolumn{1}{l|}{} &
  \textit{\textbf{correct}} &
  \textbf{\{SXAI\} \textless \{CFF\}} &
  \begin{tabular}[c]{@{}l@{}}0.08\\ (0.04)\end{tabular} &
  \begin{tabular}[c]{@{}l@{}}0.29\\ (0.04)\end{tabular} &
  \begin{tabular}[c]{@{}l@{}}$F_{1, 90.51} = 14.75$, \\ $p = .0002$, $d = .69$\end{tabular} &
  \textbf{\{SXAI\} \textless \{CFF\}} &
  \begin{tabular}[c]{@{}l@{}}0.09\\ (0.04)\end{tabular} &
  \begin{tabular}[c]{@{}l@{}}0.26\\ (0.04)\end{tabular} &
  \begin{tabular}[c]{@{}l@{}}$F_{1, 99.15} = 9.39$, \\ $p = .003$, $d = .57$\end{tabular} \\ \cline{2-10} 
\multicolumn{1}{l|}{} &
  \textit{\textbf{overrelied}} &
  \textbf{\{CFF\} \textless \{SXAI\}} &
  \begin{tabular}[c]{@{}l@{}}0.77\\ (0.05)\end{tabular} &
  \begin{tabular}[c]{@{}l@{}}0.57\\ (0.05)\end{tabular} &
  \begin{tabular}[c]{@{}l@{}}$F_{1, 71.22} = 8.80$, \\ $p = .004$, $d = .50$\end{tabular} &
 \{CFF, SXAI\} &
  \begin{tabular}[c]{@{}l@{}}0.50\\ (0.06)\end{tabular} &
  \begin{tabular}[c]{@{}l@{}}0.37\\ (0.06)\end{tabular} &
  \begin{tabular}[c]{@{}l@{}}$F_{1, 82.53} = 2.45$,\\ $n.s.$, $d = .27$\end{tabular} \\ \cline{2-10} 
\multicolumn{1}{l|}{\multirow{-3}{*}{\textbf{\begin{tabular}[c]{@{}l@{}}Carb Source\\ Detection\\ Performance\end{tabular}}}} &
  \textit{\textbf{human error}} &
  \{CFF, SXAI\} &
 \begin{tabular}[c]{@{}l@{}}0.16\\ (0.04)\end{tabular} &
  \begin{tabular}[c]{@{}l@{}}0.15\\ (0.03)\end{tabular} &
  \begin{tabular}[c]{@{}l@{}}$F_{1, 52.58} = 0.09$, \\ $n.s.$, $d = .04$\end{tabular} &
  \{CFF, SXAI\} &
  \begin{tabular}[c]{@{}l@{}}0.38\\ (0.05)\end{tabular} &
  \begin{tabular}[c]{@{}l@{}}0.39\\ (0.05)\end{tabular} &
  \begin{tabular}[c]{@{}l@{}}$F_{1, 51.75} = 0.002$, \\ $n.s.$, $d = .02$\end{tabular} \\ \Xhline{2\arrayrulewidth}
\end{tabular}
}
\caption{Distribution of decisions on instances where the model was incorrect disaggregated by NFC. Differences indicated by the \textless\ symbol are statistically significant. Categories that share brackets do not differ significantly.}
\label{incorrectNFC}
\end{table}


Participants were split into two groups --- low NFC and high NFC --- at the median of the NFC score distribution. Table~\ref{table:performance-NFC} summarizes the results of performance disaggregated by NFC. 

\paragraph{\textbf{Objective Measures.}} 
 As an initial check we compared the overall performance across categories (i.e., \emph{cognitive forcing functions} and \emph{simple explainable AI}) between the two NFC groups. High-NFC participants demonstrated significantly higher overall performance $(M = 0.44)$ than low-NFC participants $(M = 0.23)$, $(F_{1, 172.5} = 29.13, p \ll .0001)$. They also detected the source of carbs $(M = 0.67)$ significantly better $(F_{1, 172.2} = 32.11, p \ll .0001)$ than low-NFC participants $(M = 0.44)$. These results demonstrate that high-NFC participants generally perform better at this task than low-NFC participants.

Consistent with the main findings (reported in Section~\ref{section:results}), compared to \emph{simple explainable AI}, \emph{cognitive forcing functions} did not have a significant effect in overall performance and carb source detection performance on all questions for either of the groups. On incorrect model predictions, however, high-NFC participants benefited from \emph{cognitive forcing functions} as they significantly improved both their overall and carb source detection performance. Whereas, while \emph{cognitive forcing functions} improved low NFC participants' carb source detection performance significantly on incorrect model predictions, they did not significantly improve their overall performance.

\paragraph{\textbf{Detailed analysis of decisions when model predictions were incorrect.}}

For high NFC participants, consistent with the main findings, distributions of overreliance, human error, and correctness were significantly different between the CFF and SXAI categories for both the overall performance ($\chi^2(2, N=357) = 13.12$, $p = .001$) and carb source detection performance ($\chi^2(2, N=357) = 29.03$, $p \ll .0001$). As shown in Table~\ref{incorrectNFC}, \emph{cognitive forcing functions} significantly improved high NFC participants' overall and carb source detection performance compared to \emph{simple explainable AI} conditions. High-NFC participants in \emph{cognitive forcing functions} conditions, also, overrelied on the AI significantly less for carb source detection than those in \emph{simple explainable AI} conditions.

For low-NFC participants, distributions of overreliance, human error, and correctness were not significantly different between the CFF and SXAI categories for overall performance ($\chi^2(2, N=306) = 2.01$, $n.s.$). For carb source detection, however, the distributions were significantly different ($\chi^2(2, N=306) = 16.30$, $p = .0003$). As shown in Table~\ref{incorrectNFC}, participants in \emph{cognitive forcing function} conditions detected the carb source significantly better than those in \emph{simple explainable AI} conditions. Different from combined findings, there was no significant difference across categories for overreliance and human errors.

\begin{table}
\resizebox{\textwidth}{!}{
\begin{tabular}{ll|l|l|l|l|l|l|l}
 &
  \multicolumn{4}{c|}{\textbf{High NFC}} &
  \multicolumn{4}{c}{\textbf{Low NFC}} \\ \cline{2-9} 
 &
  \multicolumn{1}{c|}{} &
  \multicolumn{2}{c|}{\textbf{\begin{tabular}[c]{@{}c@{}}Marginal Means\\ (Standard Errors)\end{tabular}}} &
  \multicolumn{1}{c|}{} &
   &
  \multicolumn{2}{c|}{\textbf{\begin{tabular}[c]{@{}c@{}}Marginal Means\\ (Standard Errors)\end{tabular}}} &
  \multicolumn{1}{c}{} \\ \cline{3-4} \cline{7-8}
 &
  \multicolumn{1}{c|}{\multirow{-2}{*}{\textbf{}}} &
  \textbf{SXAI} &
  \textbf{CFF} &
  \multicolumn{1}{c|}{\multirow{-2}{*}{\textbf{Significance}}} &
  \multirow{-2}{*}{} &
  \textbf{SXAI} &
  \textbf{CFF} &
  \multicolumn{1}{c}{\multirow{-2}{*}{\textbf{Significance}}} \\ \hline
\multicolumn{1}{l|}{\textbf{Trust}} &
  \textbf{\{CFF\} \textless\ \{SXAI\}} &
  \begin{tabular}[c]{@{}l@{}}3.89\\ (0.12)\end{tabular} &
  \begin{tabular}[c]{@{}l@{}}3.56\\ (0.11)\end{tabular} &
  \begin{tabular}[c]{@{}l@{}}$F_{1, 64.22} = 5.90$, \\ $p = .02$\end{tabular} &
  \{CFF, SXAI\} &
  \begin{tabular}[c]{@{}l@{}}3.93\\ (0.13)\end{tabular} &
  \begin{tabular}[c]{@{}l@{}}3.93\\ (0.13)\end{tabular} &
  \begin{tabular}[c]{@{}l@{}}$F_{1, 86.09} = 0.0$, \\ $n.s.$\end{tabular} \\ \hline
\multicolumn{1}{l|}{\textbf{Preference}} &
  \textbf{\{CFF\} \textless\ \{SXAI\}} &
  \begin{tabular}[c]{@{}l@{}}3.75\\ (0.13)\end{tabular} &
  \begin{tabular}[c]{@{}l@{}}3.44\\ (0.12)\end{tabular} &
  \begin{tabular}[c]{@{}l@{}}$F_{1, 62.04} = 4.63$, \\ $p = .03$\end{tabular} &
  \{SXAI, CFF\} &
  \begin{tabular}[c]{@{}l@{}}3.73\\ (0.13)\end{tabular} &
  \begin{tabular}[c]{@{}l@{}}3.86\\ (0.13)\end{tabular} &
  \begin{tabular}[c]{@{}l@{}}$F_{1, 72.42} = 0.60$, \\ $n.s.$\end{tabular} \\ \hline
\multicolumn{1}{l|}{\textbf{Mental Demand}} &
  \{SXAI, CFF\} &
  \begin{tabular}[c]{@{}l@{}}2.50\\ (0.16)\end{tabular} &
  \begin{tabular}[c]{@{}l@{}}2.70\\ (0.14)\end{tabular} &
  \begin{tabular}[c]{@{}l@{}}$F_{1, 70.69} = 0.96$, \\ $n.s.$\end{tabular} &
  \{CFF, SXAI\} &
  \begin{tabular}[c]{@{}l@{}}3.27\\ (0.20)\end{tabular} &
  \begin{tabular}[c]{@{}l@{}}2.96\\ (0.19)\end{tabular} &
  \begin{tabular}[c]{@{}l@{}}$F_{1, 75.97} = 1.51$, \\ $n.s.$\end{tabular} \\ \hline
\multicolumn{1}{l|}{\textbf{System Complexity}} &
  \textbf{\{SXAI\} \textless\ \{CFF\}} &
  \begin{tabular}[c]{@{}l@{}}2.64\\ (0.15)\end{tabular} &
  \begin{tabular}[c]{@{}l@{}}2.95\\ (0.13)\end{tabular} &
  \begin{tabular}[c]{@{}l@{}}$F_{1, 57.53} = 6.26$, \\ $p = .01$\end{tabular} &
  \{SXAI, CFF\} &
   \begin{tabular}[c]{@{}l@{}}2.96\\ (0.16)\end{tabular} &
  \begin{tabular}[c]{@{}l@{}}3.11\\ (0.15)\end{tabular} &
  \begin{tabular}[c]{@{}l@{}}$F_{1, 71.73} = 0.47$, \\ $n.s.$,\end{tabular}
\end{tabular}}
\caption{Results of subjective measures disaggregated by NFC. Differences indicated by the \textless\ symbol are statistically significant. Categories that share brackets do not differ significantly.}
\label{subjective-NFC}
\end{table}

\paragraph{\textbf{Subjective Measures.}}

We first compared subjective measures reported by high NFC participants with those reported by low NFC participants by combining the data across categories (i.e., \emph{cognitive forcing functions} and \emph{simple explainable AI}). Low NFC participants generally found the task significantly more mentally demanding $(M = 3.10)$ than high NFC participants $(M = 2.62)$,  $(F_{1, 165} = 7.07, p = .009)$. Similarly, they found the system to be significantly more complex $(M = 3.04)$  than high NFC participants $(M = 2.65)$,  $(F_{1, 162.4} = 5.86, p = .02)$. Low NFC participants reported on average higher trust $(M = 3.92)$ than high NFC participants $(M = 3.70)$, albeit not significantly so $(F_{1, 172.1} = 2.89, n.s.)$. They also preferred the systems on average more $(M = 3.80)$ than high NFC participants $(M = 3.56)$, but this difference was also not statistically significant $(F_{1, 170} = 2.63, n.s.)$.

Subsequently, we investigated the effect of category on subjective ratings separately for each of the two NFC groups. These results are summarized in Table~\ref{subjective-NFC}.
High NFC participants reported significantly higher trust for \emph{simple explainable AI} conditions compared to \emph{cognitive forcing functions}. They also preferred \emph{simple explainable AI} conditions more than \emph{cognitive forcing functions} and perceived them to be significantly less complex.
There were no significant differences for subjective ratings across categories for low NFC participants.

\section{Discussion}

Consistent with prior research~\cite{bucinca20:proxy,green2019principles,lai2019human, bussone2015role}, our results demonstrate that people aided by simple explainable AI interfaces performed better overall than people who completed the task without any AI support. Also consistent with the prior research, these human+AI teams performed worse than the AI model alone (its accuracy was set to 75\%). Our results suggest that part of the reason is that people frequently overrelied on the AI: they followed the suggestions of the AI even when its predictions were incorrect. Consequently, when the AI predictions were incorrect, people aided by the simple explainable AI approaches performed less well than people who had no AI support. 

Our results demonstrate that cognitive forcing functions reduced overreliance on AI compared to the simple explainable AI approaches: When the simulated AI model was incorrect, participants disregarded the AI suggestions and made the optimal choices significantly more frequently in the \emph{cognitive forcing function} conditions than in the \emph{simple explainable AI} conditions. These results \textbf{support the hypothesis (H1a).} 

However, even with the cognitive forcing functions, the human+AI teams in our study continued to perform worse than the AI model alone: the cognitive forcing functions reduced, but not yet eliminated, overreliance on the AI.

Our results \textbf{did not provide support for the hypothesis (H1b)} that cognitive forcing functions will improve the performance of human+AI teams. There was no significant difference in performance between \emph{simple explainable AI} approaches and \emph{cognitive forcing functions}. 

Subjective ratings of the conditions indicate that participants in \emph{cognitive forcing function} conditions perceived the system as more complex than those in \emph{simple explainable AI} conditions.
The tension between subjective measures and performance is observed in the correlation analyses of the subjective measures versus the performance. When the AI model was making incorrect predictions, people performed best in conditions that they preferred and trusted the least, and that they rated as the most difficult. Thus, \textbf{as hypothesized (H2)}, there appears to be a trade-off between the acceptability of a design of the human-AI collaboration interface and the performance of the human+AI team.


Overall, our research suggests that human cognitive motivation moderates the effectiveness of explainable AI solutions. 
Hence, in addition to tuning explanation attributes such as soundness and completeness~\cite{kulesza2013too}, or faithfulness, sensitivity, and complexity~\cite{bhatt2020evaluating}, explainable AI researchers should ensure that people will exert effort to attend to those explanations. 


Our results also lend further support to the recent research that demonstrated that using proxy tasks (where participants are asked, for example, to predict the output of an algorithm based on the explanations it generated) to evaluate explainable AI systems may lead to different results than testing the systems on actual decision-making tasks (where participants are asked to make decisions aided by an algorithm and its explanations)~\cite{bucinca20:proxy}. To explain their results, the authors hypothesized that proxy tasks artificially forced participants to pay attention to the AI, but when participants were presented with actual decision-making tasks, they focused on making the decisions and allocated fewer cognitive resources to analyzing AI-generated suggestions and explanations. 
Our results showed that cognitive forcing interventions improved participants' ability to detect AI's errors in an actual decision-making task, where the human is assisted by the AI while making decisions. Hence, we concur: If we consider proxy tasks to be a strong form of a cognitive forcing intervention, our results suggest that evaluations that use proxy task are likely to produce more optimistic results (i.e., show higher human performance) than evaluations that use actual decision-making tasks with simple explainable AI approaches. 

Because there is prior evidence that cognitively-demanding user interface designs benefit people differently depending on the level of their cognitive motivation, we disaggregated our results by dividing our participants into two halves based on their Need for Cognition (NFC)---a stable personality trait that captures how much a person enjoys engaging in effortful cognitive activities.
The significant improvements in performance stemming from cognitive forcing functions, when disaggregated, held mostly for people with high NFC.
However, high-NFC participants trusted and preferred cognitive forcing functions less than simple explainable AI approaches, while there was no negative impact of these interventions on the low-NFC participants. These finding suggest that cognitive forcing functions might produce intervention-generated inequalities by exacerbating the differences in performance between people based on their NFC. These findings also point to a possible way to mitigate this undesired effect. Future interventions might be tailored to account for the differences in intrinsic cognitive motivation: stricter interventions might benefit more and still be accepted by people with lower intrinsic cognitive motivation.

A key limitation of our work is that it was conducted in the context of a single non-critical decision-making task. Additional work is needed to examine if the effects we observe generalize across domains and settings. However, because prior research provides ample evidence that even experts making critical decisions resort to heuristic thinking (and fall prey to cognitive biases)~\cite{berner2008overconfidence,bornstein2001rationality,croskerry2003importance,lambe2016dual}, we have some confidence that our results will generalize broadly.

Another limitation is that cognitive forcing functions improved the performance of the human+AI teams at the cost of reducing the perceived usability. Consequently, systems that employ cognitive forcing functions may find less adoption than those that use simple explainable AI techniques because users will resist using them. We believe that this limitation could be overcome in part by developing \emph{adaptive} strategies where the most effective cognitive forcing functions are deployed only in a small fraction of cases where using them is predicted to substantially improve the final decision made by the human+AI team. Developing such adaptive strategies will require creating efficient models for predicting the performance of human+AI teams on particular task instances in the presence of different interventions. Our work suggests that these predictive models should consider AI's uncertainty for task instances and individual differences in cognitive motivation when predicting whether cognitive forcing functions are necessary.

\section{Conclusion}

As AIs assist humans increasingly in decision-making tasks ranging from trivial to high-stakes, it is imperative for people to be able to detect incorrect AI recommendations. This goal is also an incumbent milestone towards building human and AI teams that outperform both people and AIs alone. Thus, in this study, we investigated the effect of cognitive forcing functions as interventions for reducing human overreliance on AI in collaborative human+AI decision-making.

First, our results demonstrated that cognitive forcing functions, which elicit analytical thinking, significantly reduced overreliance on AI compared to the simpler approaches of presenting explanations for the AI recommendations to the user.

Second, our results also showed that there exists a trade-off between subjective trust and preference in a system and performance with the system in human+AI decision-making. Participants preferred and trusted more the systems that they perceived as less mentally demanding, but with which they performed poorly.

Third, our research suggests that as researchers, we should audit our work for emerging inequalities among relevant groups. Our results show that cognitive forcing functions disproportionately benefited participants with high Need for Cognition --- a group that has been shown in other settings to benefit the most from useful but cognitively demanding user interface features.

Together, these findings suggest that research in explainable AI should not assume that participants will engage with the explanations by default. Hence, more effort should be spent on devising interventions that elicit analytical thinking and engagement with explanations when necessary to avoid unquestioning trust from humans. Also, further research is necessary to account for individual differences in cognitive motivation and explore the \emph{right} amount and timing of \emph{cognitive forcing} necessary for optimal human performance with AI-powered decision-support tools.

\section*{Acknowledgments}
We thank Isaac Lage, Andrzej Romanowski, Ofra Amir, Zilin Ma, and Vineet Pandey for helpful suggestions and discussions.

\bibliographystyle{ACM-Reference-Format}
\bibliography{references,kzg}
\end{document}